\documentclass[9pt,twocolumn,twoside]{osajnl}
\journal{josaa}
\setboolean{shortarticle}{false} 
\usepackage{amsmath,amsfonts,amssymb}
\usepackage{subfigure}
\usepackage{graphicx}
\usepackage{wrapfig}
\usepackage{makecell}
\usepackage{color,soul}
\usepackage[normalem]{ulem}
\long\def\comment#1{}

\def\bA{\boldsymbol{A}}
\def\ba{\boldsymbol{a}}

\def\bB{\boldsymbol{B}}

\def\bC{\boldsymbol{C}}

\def\bD{\boldsymbol{D}}
\def\bd{\boldsymbol{d}}

\def\rm{\mathrm{m}}

\def\bn{\boldsymbol{n}}

\def\bp{\boldsymbol{p}}

\def\rT{\mathrm{T}}

\def\bw{\boldsymbol{w}}

\def\bx{\boldsymbol{x}}
\def\by{\boldsymbol{y}}

\def\bphi{\boldsymbol{\phi}}

\title{Millisecond Exoplanet Imaging, I: Method and Simulation Results}

\author[1,*]{Alexander T. Rodack}
\author[2]{Richard A. Frazin}
\author[1]{Jared R. Males}
\author[1]{Olivier Guyon}

\affil[1]{Steward Observatory, University of Arizona, Tucson, AZ 85721}
\affil[2]{ Dept. of Climate and Space Sciences and Engineering, University of Michigan, Ann Arbor, MI 48109} 
\affil[*]{E-mail: atrodack@email.arizona.edu}

\dates{Compiled \today}

\ociscodes{010.1080 Adaptive Optics, 010.7350   Wavefront Sensing}

\doi{\url{http://dx.doi.org/10.1364/ao.XX.XXXXXX}}

\begin{abstract}
	
One of the top priorities in observational astronomy  is the direct imaging and characterization of extrasolar planets (exoplanets) and planetary systems.
Direct images of rocky exoplanets are of particular interest in the search for life beyond the Earth, but they tend to be rather challenging  targets since they are orders-of-magnitude dimmer than their host stars and are separated by small angular distances that are comparable to the classical $\lambda/D$ diffraction limit, even for the coming generation of 30 m class  telescopes.
Current and planned efforts for ground-based direct imaging of exoplanets combine high-order adaptive optics (AO) with a stellar coronagraph observing at wavelengths ranging from the visible to the mid-IR.
The primary barrier to achieving high contrast with current direct imaging methods is the quasi-static speckles, caused largely by non-common path aberrations (NCPA) in the coronagraph optical train.
Recent work has demonstrated that millisecond imaging, which effectively “freezes” the atmosphere’s turbulent phase screens, should allow the wavefront sensor (WFS) telemetry to be used as a probe of the optical system to measure the NCPA.
Starting with a realistic model of a telescope with an AO system and a stellar coronagraph, this article provides simulations of several closely related regression models that take advantage of millisecond telemetry from the WFS and coronagraph’s science camera.
The simplest regression model, called the na\"ive estimator, does not treat the noise and other sources of information loss in the WFS.
Despite its flaws,  in one of the simulations presented herein, the na\"ive estimator provided a useful estimate of an NCPA of $\sim$ 0.5  radian RMS ($\approx \lambda / 13$), with an accuracy of $\sim$ 0.06 radian RMS in one minute of simulated sky time on a magnitude 8 star. 
The \emph{bias-corrected estimator} generalizes the regression model to account for the noise and information loss in the  WFS.
A simulation of the bias-corrected estimator with four minutes of sky time included an NCPA of $\sim 0.05 \,$ radian RMS ($\approx \lambda / 130$) and an extended exoplanet scene.
The joint regression of the bias-corrected estimator simultaneously achieved an NCPA estimate with an accuracy of $\sim 5\times10^{-3} \,$radian RMS and an estimate of the exoplanet scene that was free of the self-subtraction artifacts typically associated with differential imaging.  The $5 \, \sigma$ contrast achieved by image of the exoplanet scene was $\sim 1.7 \times 10^{-4}$ at a distance of $3 \lambda/D$ from the star and $\sim 2.1 \times 10^{-5}$ at $10 \, \lambda/D$.
These contrast values are comparable to the very best on-sky results obtained from multi-wavelength observations that employ both angular differential imaging (ADI) and spectral differential imaging (SDI).
This comparable performance is despite the fact our simulations are quasi-monochromatic, which makes SDI impossible, nor do they have diurnal field rotation, which makes ADI impossible. 
The error covariance matrix of the joint regression shows substantial correlations in the exoplanet and NCPA estimation errors, indicating that exoplanet intensity and NCPA need to be estimated self-consistently to achieve high contrast.

\end{abstract}

\begin{document}
\maketitle
	
	\section{Introduction}
	\label{sec: Intro}
	
	The goal of directly imaging planets and planetary systems at optical and infrared wavelengths is among the top science priorities of the coming generation of so-called ``extremely large telescopes" (ELTs), which will have primary mirror diameters ranging from $D = 25$ to $D = 39 \,$m.
	Due to the small angular separations between the planets and their host stars, as well as significant differences in brightness between them, this imaging problem is undoubtedly one of great remaining challenges in astronomical optics.
	The most desired planetary targets for detection and characterization are planets capable of supporting life.
	Visible and near-IR spectroscopy of reflected starlight can reveal signs of biological activity in the planet atmospheric composition.
	Such measurements could be obtained with future large ground-based telescopes thanks to their collecting area and angular resolution, provided that the planets can be directly imaged. \cite{Guyon_habitable_ELT12}
	Directly imaging them will require ultimately achieving a contrast (i.e., planet-to-star intensity ratio) of $10^{-10}$\ or less (roughly independent of wavelength in reflected light) for angular separations that approach the $\lambda/D$ classical diffraction limit of an ELT.\cite{Stark_ExoEarthYield14, Males_DirectImaging2014, Brown_PlanetSearch15} 
	
\begin{wrapfigure}{l}{0pt}
		\includegraphics[width=.50\linewidth,clip=]{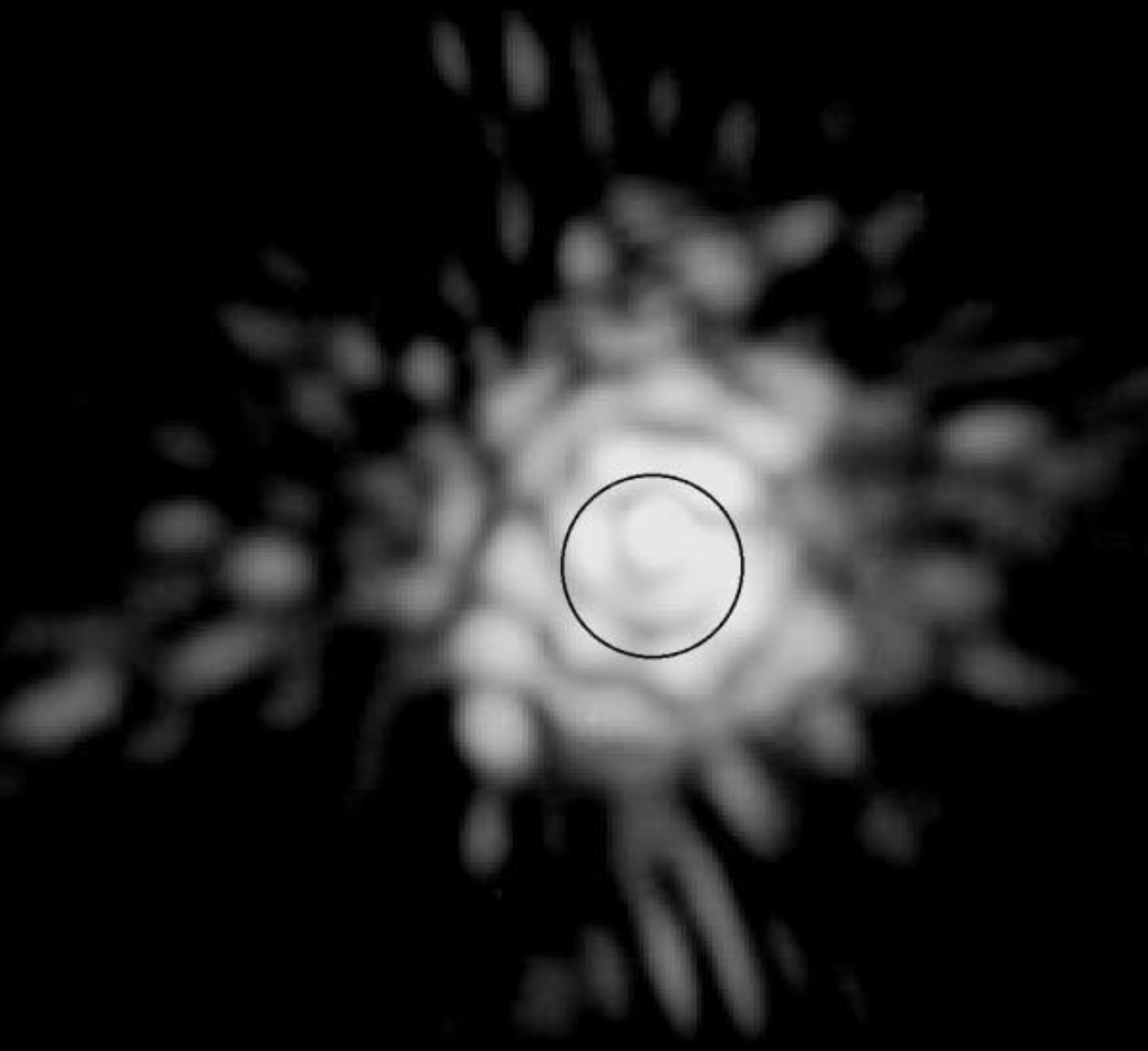}
		\caption{\small Coronagraphic AO image (logarithmic intensity) from the Lyot Project with the AEOS telescope of the bright star HD137704.  The circles indicates the coronagraphic occultation region.
			The $\sim 1200 \, \mathrm{s}$ integration time averages over the atmospheric turbulence.  $\lambda = 1.6 \, \mu \mathrm{m}$.  From [\citenum{Digby_Lyot06}].}
		\label{fig: LotsOfSpeckles}
		\vspace{-4mm}
	\end{wrapfigure}

	On the ground, imaging planets currently requires the combination of high-order adaptive optics (AO) with a stellar coronagraph \cite{Guyon_ExAOreview18}.
	As shown in Fig.~\ref{fig: LotsOfSpeckles}, the very first images acquired by early stellar coronagraphs with AO systems showed swarms of speckles, many of which were vastly brighter than any planets one might hope to see \cite{Digby_Lyot06,Racine99}.
	Achieving the high-contrast goals required to detect the desired targets in the face of these speckles remains an elusive, high-priority task decades later.
	At the heart of this speckle problem lie \emph{quasi-static aberrations (QSA)} in the optical system hardware.
	The temporal variability of the QSA results from thermal fluctuations and other variable conditions, such as the changes in gravitational stress as the telescope pointing varies.
	As a result of the ever-changing conditions, the QSA on ground-based platforms change on a wide variety of time scales, ranging from minutes to months \cite{Guyon_Limits_ApJ2005}.
	
	The most problematic of the QSA are the \emph{non-common path aberrations} (NCPA), occurring in optical components that are beyond the dichroic filter (or beam splitter) that separates the light paths that go to the AO system's wavefront sensor (WFS) and the coronagraph (see Fig.~\ref{fig: schematic}).
	Because they occur downstream of the dichroic (or beam splitter), the NCPA are either experienced by the WFS optical train, or the coronagraph optical train, but not both.
	The NCPA manifest themselves as \emph{quasi-static speckles} in the coronagraph's science camera (SC), and several authors have shown them to be a major limitation in high-contrast imaging \cite{Boccaletti04, Martinez13}.
	The quasi-static speckles are particularly problematic, because, unlike the speckles created by the atmospheric turbulence, they do not average to a spatially smooth halo that can be subtracted in post-processing in a relatively straightforward manner.
	Using lasers in the laboratory to calibrate the NCPA tends to leave large residual errors for many reasons, including differences between the laser path and path traversed by the starlight.
	For example, on a telescope with a segmented primary mirror it would be rather difficult for the laser system to capture the piston and tip/tilt errors on each segment that the starlight would experience.
	The challenge of mitigating the quasi-static speckles is made further complicated by the fact that they are modulated at kHz rates by the atmospheric aberrations that the AO system is unable to correct.

	\section{Summary of Current Techniques}
	\label{sec:currentTechniques}
	
	\subsection{Differential Imaging}
	\label{ssec:DI}
	
	Currently quasi-static speckles are removed via background subtraction techniques that employ various types of \emph{differential imaging}, the most common forms of which are \emph{Angular differential imaging (ADI)} and \emph{spectral differential imaging (SDI)}.    
	SDI takes advantage of the fact that the point-spread function (PSF) stretches with wavelength ($\lambda$), while the location of a planet does not \cite{Sparks02}.
	As this PSF stretching is proportional to the distance from the star, a weakness of this method is detecting targets at separations approaching the angular distance of $\lambda/D$, as well as targets that are extended in the radial direction, due to self-subtraction artifacts. 
	Furthermore, SDI is also limited by chromatic optical path difference \cite{Marois00, Pueyo_LOCI12, Rameau_ADI_SDI_limits15}.
	ADI takes advantage of the field rotation that occurs over the course of the night when the instrument rotator is turned off (Cassegrain focus) or adjusted to maintain fixed pupil orientation in a Nasmyth focus instrument, so that the planet appears to rotate relative to the PSF \cite{Marois06}.
	Thus, to the extent that aberrations do not evolve as the image rotates around the pointing center, correction can be achieved.  
	Some known and characterized weaknesses of ADI are similar to SDI, in that star-planet separations close to $\lambda/D$ are difficult to detect because the planet must travel an arc length of several resolution elements within the observing period \cite{Mawet_SmallAngleReview12}.
	Other effects caused by the time dependence of the planetary rotation rate around the star due to field rotation, as well as artifacts from self-subtraction, induce a host of biases that affect astrometry and photometry calculations \cite{Marois_SOSIE, Galicher_LOCI,Rameau_ADI_SDI_limits15}.
	ADI also will remove circularly symmetric components of the image, making the quality of the results dependent on the spatial arrangement of the planets and dust surrounding the target star.
	Furthermore, one of the largest issues with ADI is the fact that the quasi-static speckles change with time, which can greatly affect the subtraction accuracy.

	\begin{figure}[tbp]
		\hbox{\includegraphics[width=.99\linewidth,clip=]{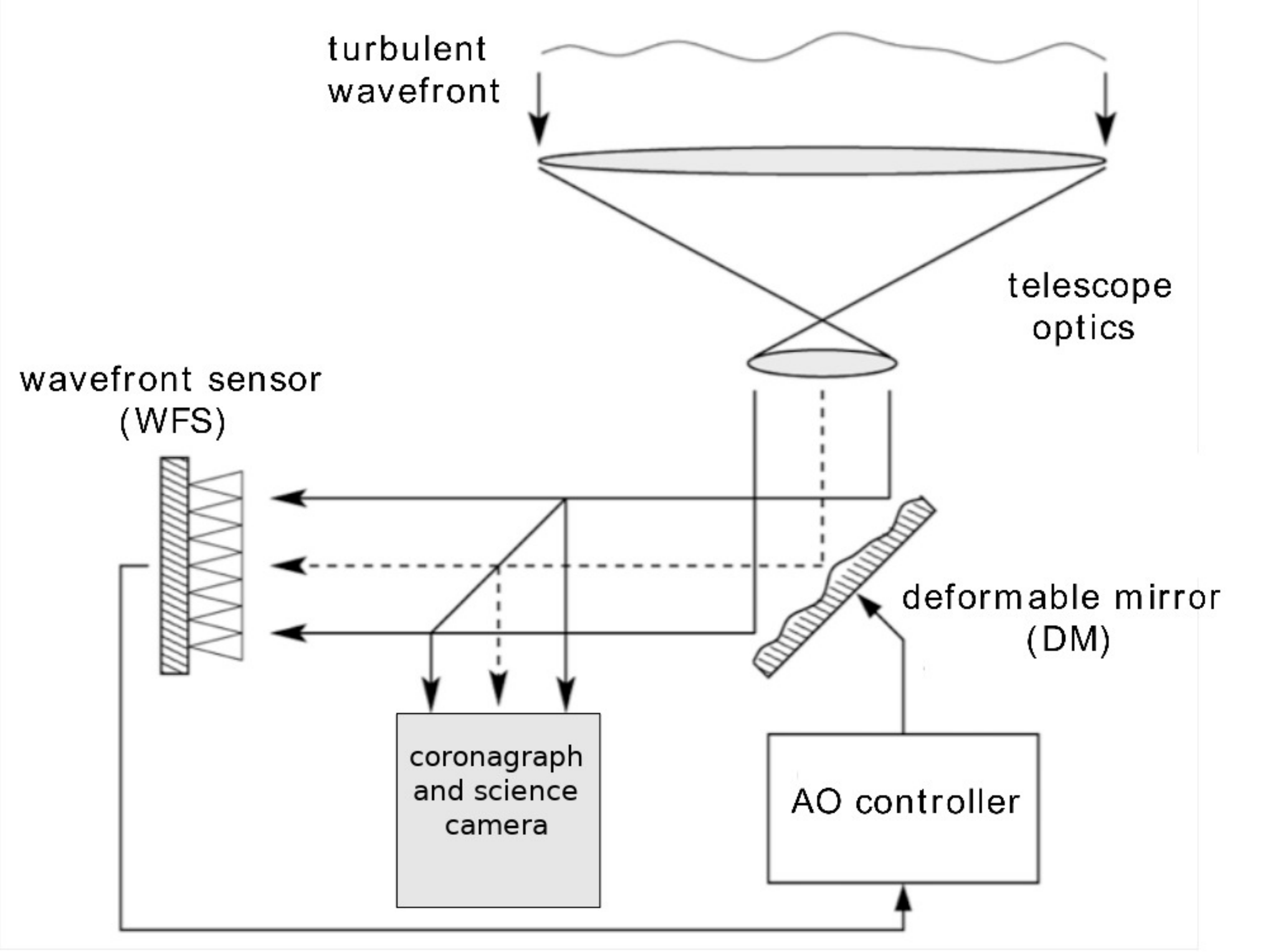}}
		\caption{\small Schematic diagram of an astronomical telescope with a closed-loop AO system and a coronagraph.  Modified from [\citenum{Hinnen_H2control}].}
		\label{fig: schematic}
		\vspace{-4mm}
	\end{figure}

	\subsection{Focal Plane Wavefront Sensing and NCPA Compensation}
	\label{ssec:FPMWS}
	
	One way to eliminate the quasi-static speckles arising from the NCPA is to perform some type of focal-plane wavefront sensing with data obtained from the science camera, and then use a deformable mirror (DM) or other type of spatial-light modulator to apply a compensation to the wavefront \cite{Martinache_SpeckleCancel14, Martinache_ClosedLoop16, Matthews_EFC_P1640_17, jovanovic2018review}.
	One way to do this is to apply two or more DM commands, called \emph{probes}, and measure the intensity for each probe, thereby setting up a simple regression in which the complex-valued electric field at the locations of interest is estimated. 
	Once this field has been estimated, the next step is to calculate a new DM command according to some merit function that creates a dark region.  
	One such method is called \emph{electric field conjugation (EFC)}, but there is now substantial literature on closely related methods \cite{Kasdin_EFC13, Martinache_SpeckleCancel14, Martinache_ClosedLoop16, Matthews_EFC_P1640_17, jovanovic2018review, Potier_2020}.
	These methods are expected to be vastly more effective on space-based platforms than they have proven to be on ground-based platforms due to the Earth’s atmospheric turbulence.
	Indeed, typical attempts to create an extended dark region on ground-based platforms only result in reducing the background starlight by a factor of about two.  
	In addition, ground-based applications of such methods are quite slow, sometimes combating a few speckles at a time, and each DM probe must be remain in place long enough to obtain a turbulent average of the intensity \cite{Thomas_GroundEFC10,Martinache_ClosedLoop16,Matthews_EFC_P1640_17}.
	The fundamental problem with ground-based probing methods and EFC (and other related methods) is that the focal plane electric field (which probing seeks to estimate and EFC seeks to cancel) is modulated by the atmosphere on the time-scale of a millisecond.
	
	A new approach to estimating the NCPA is to replace the occulter (in the first focal plane of the coronagraph) with a phase mask, thereby turning the coronagraph into a Zernike wavefront sensor \cite{Vigan_ZELDA_NCPA_AA19}.
	The Zernike wavefront sensor, combined with knowledge of the wavefront statistics and a coronagraph model, then provide sufficient information to estimate the NCPA.
	Of the various approaches to determining the NCPA, the COFFEE algorithm of Refs.~[\citenum{Paul_PhaseDivers13,Schiller_COFFEE19}], which is also referred to as ``phase diversity," is most similar to the method simulated here.
	In COFFEE, a series of DM commands (called ``pokes") is applied and a long-exposure (of sufficient duration to average over the turbulence) image is acquired.
	Then, the NCPA coefficients are estimated via regression. 
	The most important differences between method in this paper and COFFEE are:
	\begin{itemize}
		\item{Our method method uses millisecond science camera telemetry, while COFFEE uses long (i.e., averaged over the turbulence) exposures.}
		\item{Our method uses the AO residuals (available from the millisecond WFS telemetry) as probes, not DM pokes to provide the information needed to determine the NCPA.}
		\item{Our method jointly estimates the NCPA and the exoplanet image, thereby treating their joint error statistics, while the COFFEE algorithm does not estimate the exoplanet image.}
	\end{itemize}
	Note that both methods rely on coronagraph models and require knowledge of the $2\underline{nd}$ order spatial statistics of the AO residual wavefronts.

	\subsection{Millisecond Exposures}
	\label{ssec:msExposures}

\begin{wrapfigure}{l}{0pt}
		\includegraphics[width=.5\linewidth,clip=]{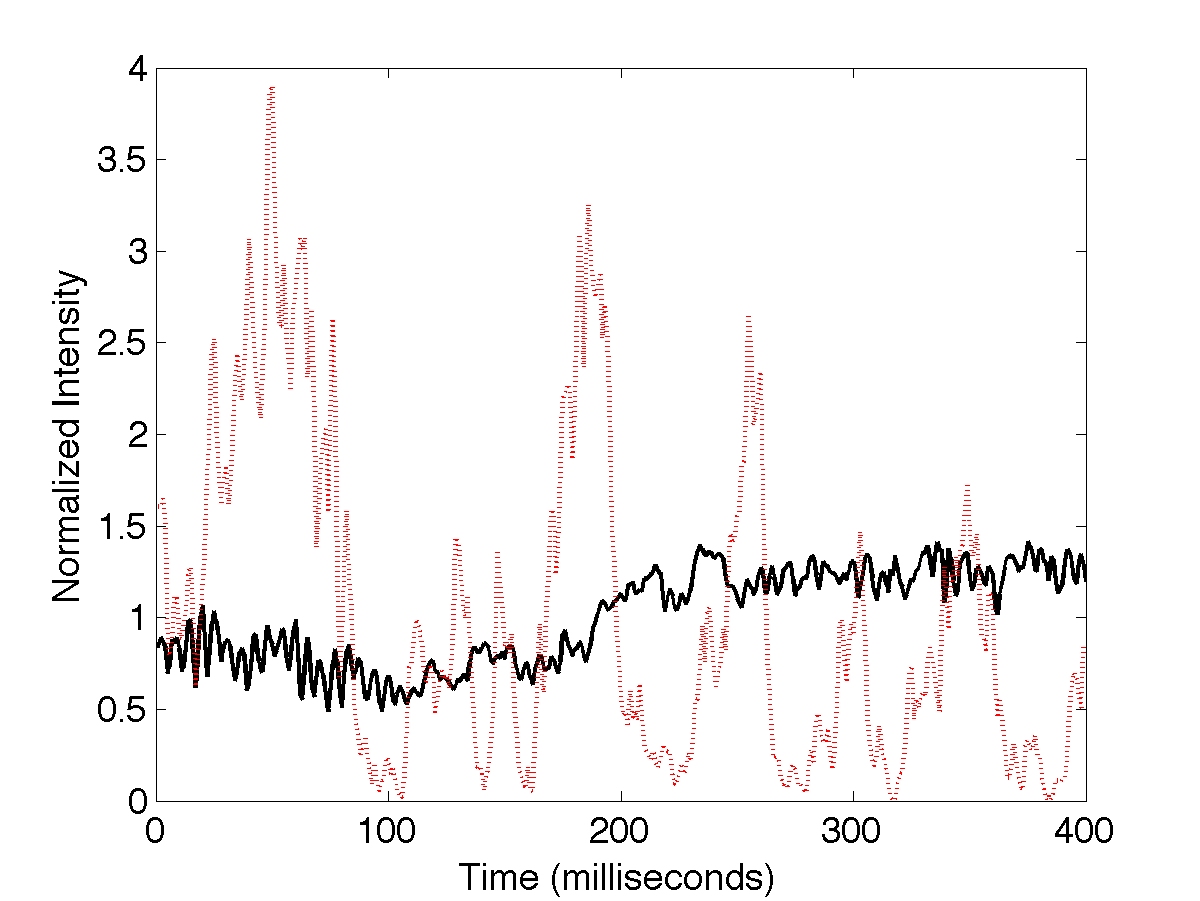}
		\caption{\small Simulation of turbulent modulation of intensities at a single pixel of the science camera from a stellar coronagraph.  The black solid line shows the time-series of the temporal variation of the planetary intensity.  The dotted red line shows the stellar intensity at the planet's location.  Both the planetary and stellar intensity are normalized to have a mean of unity in this figure. From [\citenum{Frazin13}].}
		\label{fig: modulation}
		\vspace{-5mm}
		\hspace{-10mm}
	\end{wrapfigure}

	Millisecond exposure times with the science camera are a new frontier in high contrast imaging, and they are becoming observationally attractive due to a new generation of noiseless and nearly noiseless IR and near-IR detector arrays capable of millisecond read-out times \cite{Saphira_eAPD14,Mazin_MKIDS18}.
	Millisecond exposures freeze the swarms of speckles that arise due to atmospheric turbulence, whereas longer exposures average these speckles into a smooth halo.  
	A planet and the stellar speckle exhibit radically different behavior at millisecond time-scales, as shown in Fig.~\ref{fig: modulation} \cite{Frazin13}.
	This is because at the center of a planet's speckle pattern, the AO system maintains the planet's intensity at a nearly constant level (given by the Strehl ratio), while the starlight at the planet's location exhibits much more volatility \cite{Fitz_SpeckleStats06,Gladysz10}.
	Several methods have been proposed to exploit this difference in temporal behavior based on millisecond science camera time-series data alone \cite{Stangalini_RecurQuant18,Walter_ModifiedRician19}.
	
\begin{figure*}[htbp]
		\centering
		\hbox{\includegraphics[width=.99\linewidth,clip=]{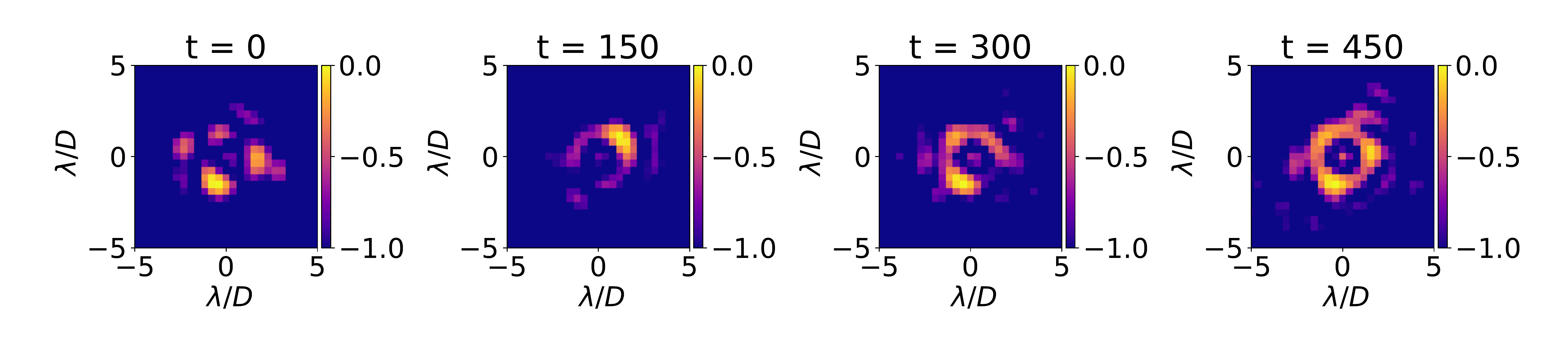}}
		\caption[short]{\small Log10 scale, noise-free focal plane images for different realizations of the instantaneous AO residual phases, showing the modulation of the signal. The range of the color scales are limited to a factor of 10 to make the modulation easy to see.
		}
		\label{Fig: NCPA coupling}
	\end{figure*}

	Simple physical optics arguments show that the speckles caused by the NCPA are modulated by the AO residual at the kHz time scale,\cite{Sauvage_model10} and Frazin showed that knowing the values of the AO residuals allows joint estimation of the NCPA and the exoplanet image from the millisecond science camera images \cite{Frazin13}.
	Fig.~\ref{Fig: NCPA coupling} shows simulations of noise-free science camera images with the same NCPA being probed with different AO residual wavefronts demonstrating this phenomenon.
	In 2013, independent publications by Frazin and Codona \& Kenworthy proposed to exploit the WFS telemetry in addition to the millisecond science camera images for focal-plane wavefront sensing \cite{Codona13, Frazin13}.
	The fundamental shortcoming of the methods of Frazin and Codona \& Kenworthy is that they were unable to account for wavefront measurement error (WME).
	This is to say, the WFS measurements only allow an imperfect estimate of the phase of the wavefront impinging on the WFS entrance pupil, not the actual phase.
	Specifically, any WFS exhibits spatial and temporal bandwidth limitations (the latter are less important due to the kHz frame-rate), nonlinearity in the phase of the wavefront, and noise.
	Outside of the high-contrast imaging problem, several authors have performed multi-frame deconvolution with millisecond focal plane and WFS telemetry to remove atmospheric image distortion.
	These works use the WFS-based estimates of the point-spread function (PSF) as an initial guess \cite{Fusco_AOdecon99, Jefferies_WFSdecon13}.
	(For approaches that do not make use of WFS telemetry see, e.g., Refs.~[\citenum{Schulz_HubbleBDC97, Fetick_BDC20}]).
	The regression results shown in this article are the first to employ a statistically rigorous treatment of the wavefront measurement error (WME), and details of the regression model are deferred to Part II due to their complexity.
	
	\section{Numerical Simulation}
	\label{sec:NumericalModels}
	
	In order to proceed with the verification of the regression model described in Part II, a suite of simulation tools was developed in order to run numerical experiments using a millisecond imaging, adaptive optics coronagraph.
	An end-to-end simulator was constructed in order to create turbulent wavefronts, as well as models for the coronagraph and the AO system optical trains.
	The AO system includes a deformable mirror (DM) followed by a 4-sided pyramid WFS (PyWFS), and the coronagraph is a standard Lyot model, followed by a science camera.
	These simulations only require pupil planes and focal planes, and the propagation between the two planes is computed via fast Fourier transforms \cite{IntroFourierOptics}.

	\subsection{Wavefronts and AO}
	\label{ssec:WavefrontsandWFS}

	Fig.~\ref{fig: schematic} shows a schematic diagram of a telescope with an AO system and a coronagraph.
	The turbulent wavefront at the telescope entrance pupil is taken to be at the science wavelength, $\lambda$ (chromatic issues will need to be considered in subsequent studies; a choice made to simplify the discussion in this work).
	We represent the entrance pupil of the telescope as comprised of $P$ pixels, with the phase at the $p$\underline{th} pixel at time $t$ represented as $\phi_p(\bw_{t}^- )$, where the $^-$\ superscript indicates that the wavefront has not yet impinged on the DM and only carries the effects of atmospheric turbulence (the post-DM wavefront will not have the superscript).  
	The function $ \phi_p $ and the parameter vector $\bw_t$ together specify the wavefront.
	There are many ways to do this.  
	For example, $\phi_p(\bw_t)$ could represent a sum over Zernike modes (evaluated at the location of pixel $p$), in which case the vector $\bw_t$ would be a set of coefficients of the  Zernike modes at time $t$.
	Here we make a simpler choice: the $p\underline{th}$ component of $w_t$ is the phase at pixel $p$ and the $\phi_p$ function just returns the value of the $p\underline{th}$ component of the input vector.
	This was called the \emph{PhasePixel} representation in Ref.~\cite{Frazin_JOSAA2018}, where some of its computational advantages are explained.
	With this representation, the complex-valued electric field in the pre-DM pupil plane is
	\begin{equation}
	u_p^-(\bw_t^-) = \exp[ j \phi_p(\bw_{t}^- )  ] \, ,
	\label{eq: Telescope entrance field}
	\end{equation}
	where $t$ is the time index in the range $0$ to $T-1$, with $T$ being the total number of millisecond exposures.
	
	The temporal evolution of the on-sky turbulent wavefronts must be specified in order to model the functionality of the AO system.
	We have taken the spatial statistics of the turbulence to be given by the von K\'arm\'an spectrum, with an outer scale parameter $L_0$, inner scale parameter $l_0$, and Fried parameter $r_0$ \cite{StatisticalOptics}.
	To keep the model simple, and maintain realism in the temporal evolution, we implement the infinite phase screen method introduced by Assémat et al \cite{Assemat_Wilson_InfinitePhaseScreen06} via the hcipy package \cite{Por_Haffert_Hcipy2018}.
	
	As our model does not include the telescope’s beam reducing optics, the first surface the atmospheric wavefront encounters is that of the DM.
	We simulate the DM surface with 2D cubic splines, with the knots placed in a $20 \times 20 $ grid.
	In this way, the knots play the role of the actuators and the height of any point on the surface is given by the spline interpolator function.
	With the circular pupil inscribed on the DM surface, the resulting control radius is $10 \, \lambda/D$.
	The phase imparted by the DM is added to the pre-DM phase, $\phi_p(\bw_t^-)$:
	\begin{equation}
	\phi_p(\bw_t) = \phi_p(\bw_t^-) + \phi^\mathrm{DM}_{p,t} \, ,
	\label{eq: post DM phase}
	\end{equation}
	where $\phi^\mathrm{DM}_{p,t}$ is the contribution of the DM to the phase at pixel $p$ and time $t$.  
	The $P \times 1$ vector containing all $P$ values of $\phi_p(\bw_t)$, $\bphi(\bw_t)$ , is called the \emph{AO residual wavefront}, however, we will also call $\bw_t$ by the same name for convenience. 
	The $P$ equations shown in \eqref{eq: post DM phase} define the parameter vector $\bw_t$, implicitly assuming that they can be inverted, which is a trivial matter for the PhasePixel representation used here.
	More complicated representations may require a fitting step to find $\bw_t$.  
	
	Following the DM, a beam splitter separates the coronagraph and wavefront sensor optical trains.
	In our simulations, the beam splitter's only effect is to divide the available photon flux between the optical trains.
	Moving down the AO system, the WFS telemetry then makes it possible to get an estimate (or measurement) of $\bw_t$, denoted as $\hat{\bw_t}$, with the ``hat" $\hat{\bw_t}$ designating that it is an estimate of the vector of quantities that specify the true phases $\bw_t$.
	The wavefront measurement (determined by analysis of the WFS signal) can be modeled by the equation
	\begin{equation}
	\hat{\bw}_t = \mathcal{W}(\bw_t) + \bn_t \, ,
	\label{eq: WFS measurement}
	\end{equation}
	where $\mathcal{W}$ is a nonlinear measurement operator (even if the estimator itself is linear, as it is in these simulations, the WFS signal is nonlinear in wavefront phase values), and $\bn_t$ is the noise vector.
	The nonlinearity of $\mathcal{W}$, aliasing, and finite bandwidth effects tend to make $\hat{\bw}$ a biased estimate of $\bw$.
	
\begin{figure}[htbp]
		\centering
		\hbox{\includegraphics[width=.99\linewidth,clip=]{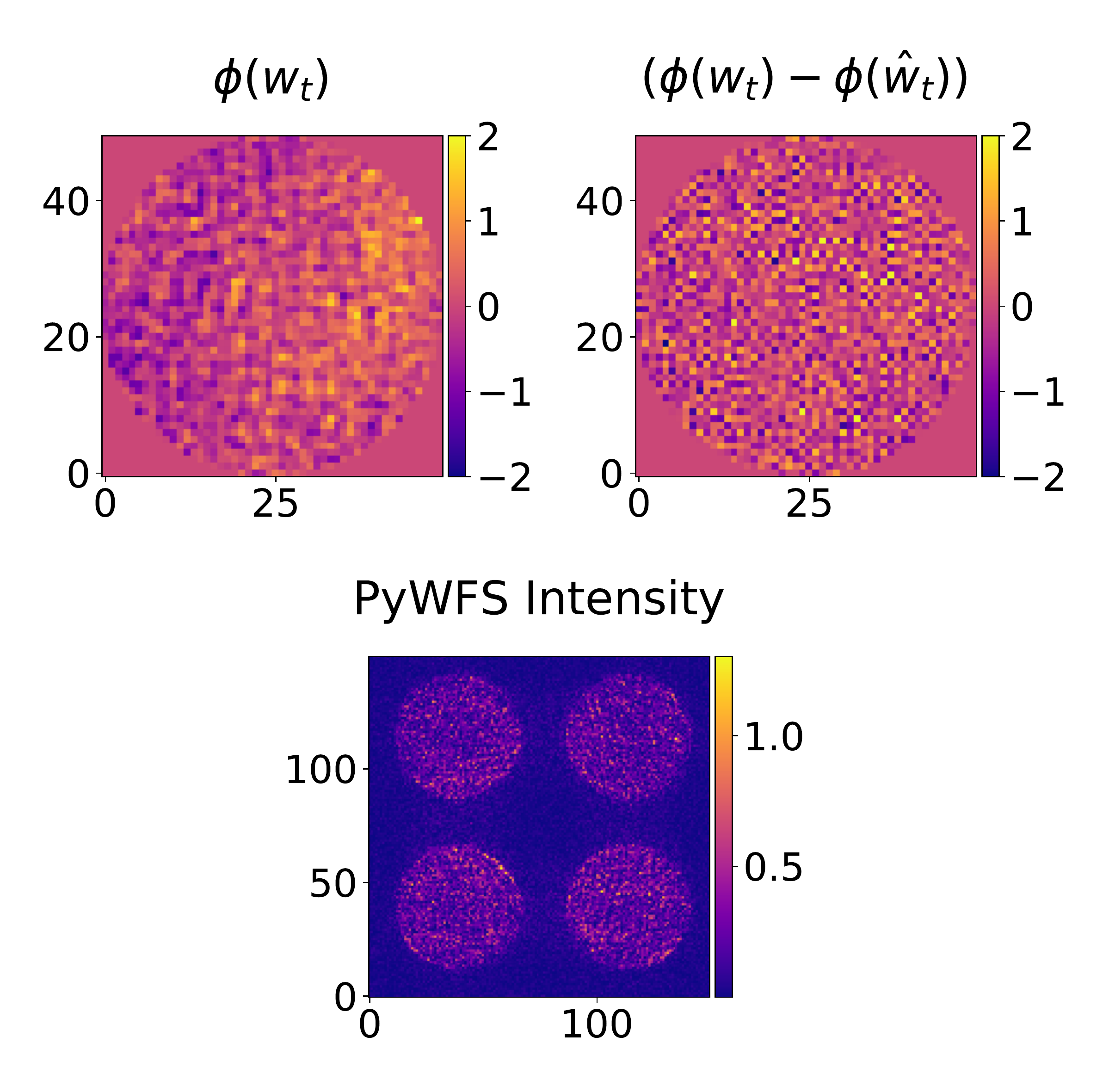}}
		\caption[short]{\small Example of wavefront measurement with the simulated PyWFS.  \emph{top left:} True AO residual (radian).  \emph{top right:} Error in measured AO residual (radian).  \emph{bottom:}  Intensity at PyWFS detector, in normalized units.
		}
		\label{Fig: Example wavefront and measurement}
	\end{figure}

	The regression model, which jointly estimates the NCPA and exoplanet image, makes essentially no assumptions about the WFS architecture, allowing tremendous flexibility.
	In principle, any information lost in the WFS is accounted for in the Monte Carlo calculations (see below).
	In the simulation experiments below, the PyWFS is chosen in large part because it is being used for many modern AO systems including MagAO-X \cite{2018SPIE10703E..09M} and SCExAO \cite{Lozi_SCExAO_2018}.
	Ref. [\citenum{Frazin_JOSAA2018}] provides linear and nonlinear regression models for a PyWFS, and for this work we apply the linear regression model.  
	The input to the PyWFS regression model is the AO residual wavefront at time-step $t$, represented as $\bw_t$, while the output is the estimate of the same, represented as $\hat{\bw}_t$, as summarized in \eqref{eq: WFS measurement}.
	The PyWFS model includes circular modulation of the beam, simulated as a discrete set of tips and tilts.
	This formulation allows for the effects of noise, spatial bandwidth, and aliasing to be included in the measurement of the wavefront.
	An example of an AO residual phase, $\phi(\bw_t)$, its measurement error, $\phi(\bw_t) - \phi(\hat{\bw}_t)$, and the corresponding PyWFS intensity at $t$, are provided in Fig. \ref{Fig: Example wavefront and measurement}.

	Progressing to the coronagraph from the beam splitter, the wavefront that arrives at the entrance pupil of the coronagraph is the same as the wavefront at the WFS, but with the addition of the NCPA phase.
	For simplicity, the cumulative sum of any NCPA is treated as a single phase error in the coronagraph entrance pupil.
	This can be thought of as the total phase error induced by the individual optical elements in the coronagraph, aberrations upstream of the beam splitter that are not corrected by the AO system (e.g., island modes), as well as aberrations in the WFS optical train that are flattened by the AO system (for which the inverse of would appear in the coronagraph optical train).
	The electric field in the coronagraph entrance pupil is thus given by
	\begin{equation}
	u_p(\bw_t, \ba) = \exp  [ j \phi_p(\bw_t) + j \theta_p  ( \ba) ] \, ,
	\label{eq: coronagraph entrance field}
	\end{equation}
	where $\ba$ is the column vector (with $N_a$ components) containing the coefficients that specify the NCPA, and $\theta_p$ is the function that specifies the phase of the NCPA in the $p\underline{th}$ pixel in the pupil.
	In these simulations,  in which aberrations are represented by Zernike polynomials, $\theta_p(\ba)$ represents the weighted sum of Zernike polynomials evaluated at pixel $p$.

	\subsection{Coronagraph}
	\label{ssec:Coronagraph}

	\begin{figure}[htbp]
		\hbox{\includegraphics[width=.95\linewidth,clip=]{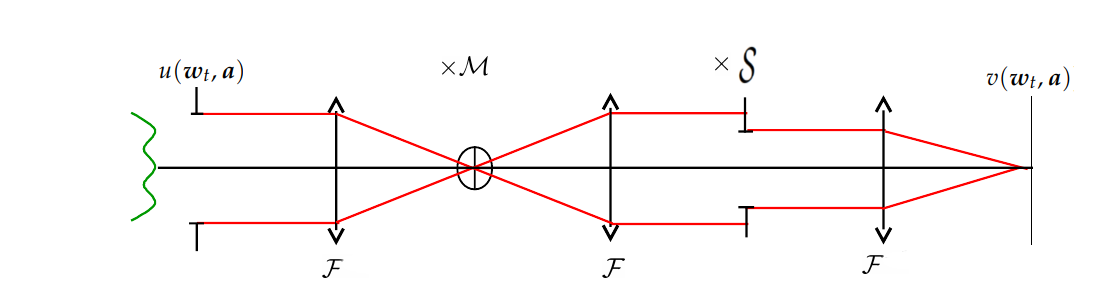}}
		\caption{\small Schematic diagram of a Lyot coronagraph.  Modified from [\citenum{Herscovici-Schiller2017}].}
		\label{fig: Lyot schematic}
		\vspace{-2mm}
	\end{figure}
	
	The regression algorithm simulated below makes no assumptions about the type of coronagraph architecture that is in the optical system, and thus provides for the freedom of many choices, including Lyot, PIAACMC \cite{Guyon_LyotPIAA14}, APP \cite{Por_APPdesign_2017,Otten_vAPPCperformance14}, and vector vortex \cite{Mawet_VectorVortex2009} coronagraphs, to suppress host starlight.
	Indeed, a coronagraph is not a mathematical necessity for running the algorithm - although one would expect coronagraphic optical systems to outperform non-coronagraphic ones.

	\begin{wrapfigure}{l}{0pt}
		\includegraphics[width=.5\linewidth,clip=]{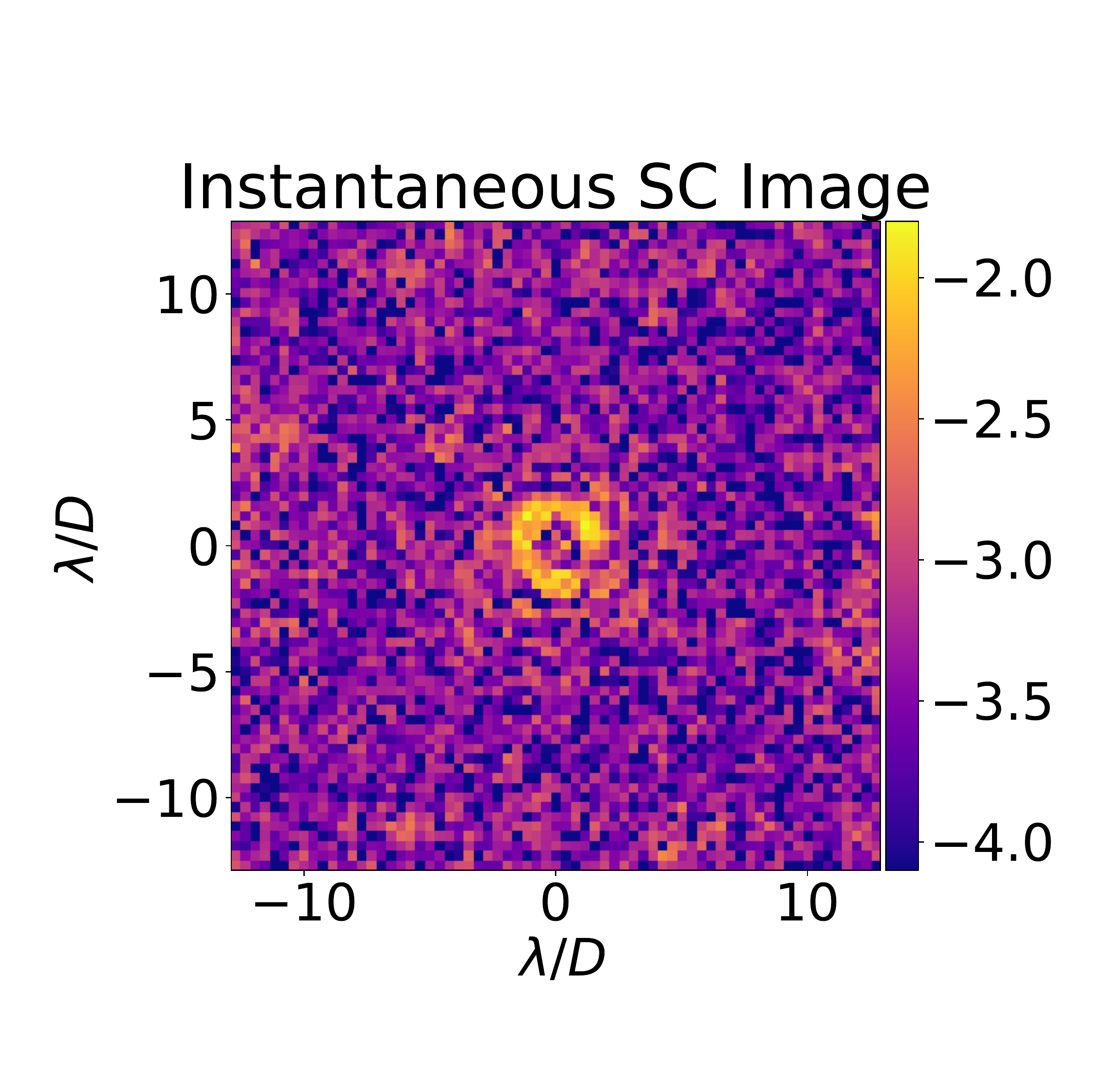}
		\caption{\small Log scale $1 \,$ms exposure science camera image in contrast units of an 8 magnitude source at $\lambda=1$ micron with $10\%$ spectral bandwidth.}
		\label{fig: noisyinstantSC}
		\vspace{-5mm}
		\hspace{-10mm}
	\end{wrapfigure}
	
	For these simulations we choose a Lyot coronagraph (See Fig.~\ref{fig: Lyot schematic}) with a binary focal plane mask (FPM) occulter that is circular in shape with a radius of $1.5 \, \lambda / D$, and a Lyot Stop with a diameter 90\% of the pupil diameter.
	The complex-valued electric field at the $l$\underline{th} science camera detector pixel, $v_l$, can be represented in terms of the complex-valued electric field in the coronagraph entrance pupil, $ u_p(\bw_t, \ba) $, via the linear coronagraph operator, a $L \times P$ matrix, $\bD = \{ D_{lp} \}$ ($L$ is the number of science camera pixels, $P$\ is the number of pixels in the discretization of the entrance pupil):
	\begin{equation}
	v_l(\bw_t, \ba) = \sum_{p=0}^{P-1} D_{lp} u_p(\bw_t, \ba)  \, ,
	\label{eq: Coronagraph Detector Field}
	\end{equation}
	where $u_p(\bw_t, \ba)$ is given by \eqref{eq: coronagraph entrance field}.
	Note that \eqref{eq: Coronagraph Detector Field} is the form of a matrix-vector multiplication and is ideally suited to being carried out on a graphics processing unit (GPU).
	$\bD$ comes from a computational model of the coronagraph optical  train.
	For a Lyot coronagraph, such a model is provided by the operator:
	\begin{equation}
	\bD = \{ \mathcal{F} \{ \mathcal{S}  \times \  \mathcal{F}[ \mathcal{M}\  \times \ \mathcal{F \big(} \:  \: \mathcal{\big)}] \} \} \, ,
	\label{eq: Coronagraph Operator}
	\end{equation}
	where $\mathcal{F}$ is the 2D discrete Fourier Transform operator, $\mathcal{S}$ is a mask representing the Lyot Stop, and $\mathcal{M}$ is a focal plane mask.
	Finally, the noise-free (i.e. true) representation of the intensity impinging on the $l\underline{th}$ pixel of the science camera is given by
	
	\begin{equation}
	I_l(\bw_t, \ba) = v_l (\bw_t, \ba) v^*_l (\bw_t, \ba) \, ,
	\label{eq: SC Intensity}
	\end{equation}
	where $^{*}$ denotes the complex conjugate.

	The regression approach allows the intensity $I_l(\bw_t, \ba)$ in \eqref{eq: SC Intensity} to be nonlinear in $\bw_t$, but it does require linearization in $\ba$ (which can be refined iteratively).  The linearization about the point $\ba_0$ is represented as:
	\begin{equation}
	I_{l}(\bw_t, \ba) \approx d_l(\bw_t) + \bA_l(\bw_t) \ba \, 
	\label{eq: SC linearization}
	\end{equation}
	where $d_l(\bw_t) = I_{l}(\bw_t, \ba_0) $ and $\bA_l(\bw_t)$ is a $1 \times N_a$ row vector representing the gradient
	\begin{equation}
	\bA_l(\bw_t) = \frac{\partial I_{l}(\bw_t, \ba)   }{\partial \ba} \bigg|_{\ba_0}  \, .
	\label{eq: row gradient}
	\end{equation}
	With this notation, $ \bA_l(\bw_t) \ba $ is a scalar product between the row vector $ \bA_l(\bw_t) $ and the column vector $\ba$.
	The $L$ values of $d_l(\bw_t)$, one for each detector pixel, are concatenated into the column vector $\bd(\bw_t)$ and, similarly, the $L$ row vectors $\bA_l(\bw_t)$ are stacked to form the 
	$L \times N_a$ matrix $\bA(\bw_t)$.
	$\bA(\bw_t)$ is called the \emph{system matrix}, and $\bd$ is called  the \emph{zero point vector}.
	The simplest situation is when the linearization point $\ba_0$ is all zeros, in which case $\bd$ corresponds to the science camera intensity without any NCPA present.

	This intensity is normalized such that a point source propagated to the final focal plane without a coronagraph has a maximum value of unity, centered at the origin of the science camera.
	This gives the effective units of the science camera measurements to be in ``contrast" as defined as the ratio of coronagraphic intensity from our simulated source divided by the non-coronagraphic intensity of an on-axis point source.
	Scaling in this fashion allows for direct calculation of contrast values.
	
	Finally, we simulate the science camera's measurement of the intensity by adding photon counting (shot) noise and detector readout noise.
	An example of such a measurement can be found in Fig.~\ref{fig: noisyinstantSC}.
	The number of photons incident on the detector is determined by choice of target star magnitude and science wavelength, optical system beam splitter ratio, and chosen values for system throughput, which are specified in Table (\ref{Table: Simulation Parameters}).
	The intensity, $I_l(\bw_t, \ba) $, in units of contrast, is converted to units of photons.
	$\sigma_{RN}$ is then chosen to represent the readout noise in photon counts per pixel, per read out (every 1 millisecond).
	The standard deviation of the intensity, in photon units, in the science camera measurement, is thus given as
	
	\begin{equation}
	\sigma_{l,ph} = \sqrt{I_{l,ph} + \sigma_{RN}^2} \, ,
	\label{eq: SC photon variance}
	\end{equation}
	allowing for the calculation of the noise term in the $l\underline{th}$ pixel via sampling from a zero-mean, $\sigma_{l_{ph}}$ Normal distribution.
	This noise term, $n_l$ , separate from the noise term in the WFS measurement in Eq. (\ref{eq: WFS measurement}) because they come from different detectors, is then converted back to units of contrast and added to the result of Eq. (\ref{eq: SC Intensity}), giving the final, measured intensity in the $l\underline{th}$ pixel, again in units of contrast, as
	\begin{equation}
	y_l(\bw_t,\ba) = I_{l}(\bw_t, \ba) + n_{l} \, .
	\label{eq: SC noisy measurement}
	\end{equation}
	To apply the regression methods, we use linearization given in \eqref{eq: SC linearization}, and we stack the $L$ SC measurements given by \eqref{eq: SC noisy measurement} into a column vector $\by_t$:
	\begin{equation}
	\by_t = \bA(\bw_t) \ba + \bd(\bw_t)  + \bn_t \, ,
	\label{eq: y w/o planets}
	\end{equation}
	where $\bn_t$ is vector that represents noise in the SC measurements at time $t$.
	When we consider $T$ exposures and allow $t$ to be an index from 0 to $T-1$,
	Eq.~\ref{eq: y w/o planets} allows one to estimate $\ba$, the NCPA coefficients.
	Further, as shown in Part II, the system matrix can be extended to include $\bp$, the $N_p \times 1$ column vector coefficients corresponding to the image of the exoplanetary system, which we will call the \emph{exoplanet image} for convenience.  When we include the exoplanet image in \eqref{eq: y w/o planets}, it takes the form:
\begin{equation}
\by_t = \bA(\bw_t) \bx + \bd(\bw_t)  + \bn_t \, , \: \: \mathrm{where} \: \: 
\bx = \left(\begin{array}{c} \ba \\ \bp \end{array} \right)
\, .
\label{eq: y}
\end{equation}
Note that $\bA(\bw_t)$ is different in Eqs.~(\ref{eq: y w/o planets}) and (\ref{eq: y}); in the former $\bA(\bw_t)$ is $L \times N_a$, whereas in the latter it is $L \times (N_a + N_p)$.
It is the latter form that allows the joint estimation of the NCPA and exoplanet brightness coefficients.
The details of constructing the system matrix can be found in the Part II article.

	\subsection{Regression Approach}
\label{ssec:RegressionApproach}

In this article, simulated kHz measurements from the science camera and WFS are used as inputs into a regression model based on Eq.~\ref{eq: y} to jointly estimate the NCPA and the extended image of circumstellar material (including planets).  
We define estimation to mean the solving Eq.~\ref{eq: y} for $\hat{\bx}$, an estimate of the true vector $\bx$, which consists of the $N_a$ NCPA basis function coefficients discussed above, as well as the $N_p$ exoplanet brightness coefficients.

This article provides simulations of the estimation methods defined in Part II:
\begin{itemize}
	\item Ideal estimation
	\item Na\"ive estimation
	\item Bias-corrected estimation
\end{itemize} 
In \emph{ideal} estimation, the true AO residuals ($\bw_t$) are used in the regression equations, resulting in an unbiased estimator (this is the method of Frazin \cite{Frazin13}).
In the real world, ideal estimation is not possible because only a measurement of the wavefront, ($\hat{\bw_t}$), can be known.
The ideal estimate will be used as a benchmark.

In \emph{na\"ive} estimation, the measurements of the AO residuals are simply plugged into the regression equations wherever they call for the true values of the AO residuals.
The fact that estimates of the AO residuals are not the same as the true AO residuals is ignored in na\"ive estimation.
It is important to understand that even if the wavefront measurements are unbiased, meaning that the expected value of the wavefront measurement error is zero, the na\"ive estimator is still biased for reasons that are explained in Part II.
In our simulations, it turns out that na\"ive estimation works rather well for estimating large ($0.5$+ radian RMS, or $\approx \lambda / 13$) NCPA, but fails to produce precise results when estimating small ($<0.1$ radian RMS, or $\approx \lambda / 62$) NCPA or the exoplanet image, as will be seen below.

Unlike the na\"ive estimate, the \emph{bias-corrected} estimate treats the wavefront measurement error (WME) under the assumption that the spatial statistics of the AO residual wavefronts are known.
This statistical knowledge is used to create a set of Monte Carlo wavefronts that form the basis of the technique.
To the extent that the Monte Carlo wavefronts are representative of the random process that generates the true AO residual wavefronts, the technique converges to an unbiased estimate with a large sample size.
In other words, when the knowledge of the statistics is exact, the bias-corrected estimator is unbiased (and nearly as good as the ideal estimate).
If the knowledge of the statistics is ``good" but not exact, the bias-corrected estimator, as is the case for the simulations shown below, retains a small bias.
The details of all three estimators are given in Part II.

The simulations presented herein are divided into Phase A and Phase B.
The Phase A simulations are designed to showcase the potential for correcting large NCPA via the na\"ive estimate with only one minute of observed data.
Thus, in Phase A, the regression model only includes the NCPA, as per \eqref{eq: y w/o planets}, and requires no knowledge of the statistics of the wavefronts.
The Phase B simulations are based on \eqref{eq: y}, and are designed to demonstrate the full potential for simultaneously estimating small NCPA and the exoplanet image.

	\subsection{Simulation Parameters}
	\label{ssec:SimulationParameters}

	In the experiments described below, the sensing and science wavelengths are both chosen to be Y band, centered around $1.036 \, \mu$m.
	The Y band was essentially an arbitrary choice corresponding to a possible MagAO-X instrument observing case; the methods discussed would work at shorter or longer wavelengths.
	Although modern ExAO systems typically use separate wavelength bands for the WFS and coronagraph optical trains, we chose to use the same band to match the current expectation for a first on-sky implementation of this method.
	This mitigates chromatic effects that may arise when these functions are performed at different wavelengths.
	A 6.5 meter, circularly symmetric aperture with no central obscuration nor spiders is chosen to define the telescope pupil.
	The light is taken to be quasi-monochromatic.
	The photon fluxes, which are needed for noise statistics, calculated from an apparent magnitude 6 or 8 sources with a 10\% spectral bandwidth and a 50\% overall throughput.
	This results in a total of 28,700 photons/ms for the magnitude 8 source and 182,200 photons/ms for the magnitude 6 source.
	70\% of the photon flux was apportioned to the WFS, while 30\% was apportioned to the coronagraph.
	This non-standard way of splitting the photons was chosen so that changing from a magnitude 6 to magnitude 8 target in the simulations would not incur a penalty in the resulting Strehl ratio from the AO system.
	We did not run experiments with other ratios.
	
	The AO system uses a leaky integrator control system with a loop gain of $0.35$, with a two frame correction delay, and provides a closed-loop Strehl ratio of $\sim 0.73$ in Y band.
	The turbulence parameters are specified in Table~\ref{Table: Simulation Parameters}.
	The control radius of the AO system (defined by the maximum correctable spatial frequency by the DM) is $10 \lambda / D$ in both the x and y directions, leading to a square corrected region, and a residual bright halo outside.
	A $1$ kHz frequency is chosen for both the wavefront and the science camera measurements, and they are assumed to collect simultaneous telemetry, meaning that $\by_t$, the vector of science camera intensities at all $L$ pixels measured at time-step $t$, results from the wavefront $\bw_t$, with measurement $\hat{\bw_t}$.
	These choices are representative of modern ground-based AO systems equipped with high frame rate science cameras.
	This AO performance, and the resulting raw contrast provided by the Lyot Coronagraph without any NCPA present, is consistent with the current theoretical performance of such a simple integrator control system given by Males and Guyon \cite{MalesandGuyon2018}.
	
	The turbulent wavefront phase, $\bphi^(\bw_t^-)$, is generated following Section\ \ref{sec:NumericalModels}.\ref{ssec:WavefrontsandWFS} and corresponds to a $50 \times 50$ pixel grid inscribed with a circular aperture with a radius of 25 pixels.
	The total number of pixels in this inscribed circle is $P = 1976$.
	The turbulence model used in the simulation is a phase only atmosphere, with $L_0=25 \,$m and $r_0=0.3 \,$m.
	The phase $\bphi(\bw_t)$ is constructed as the sum of $6$ layers (as scintillation effects are ignored to simplify the discussion in this work), each with their own wind direction vector, corresponding to a $10$ m/s speed.
	The choice of $6$ layers collapsed together in the entrance pupil helps ensure that each phase screen is unique.
	
	The science camera (in the final focal plane of the coronagraph) consists of a $67 \times 67$ pixel square array, with $0.39$ pixels per $\lambda / D$ sampling.
	A value of $0.3$ ph/pix/read of readout noise is assumed.
	This number of pixels in the detector frame, as well as the readout noise value, running at a 1kHz frame rate, is within specifications for modern, commercially available EMCCD cameras running at an acceptable gain setting.
	Table (\ref{Table: Simulation Parameters}) summarizes the general parameters that are used in the numerical experiments.
	
	The NCPA for the simulated experiments is chosen to be the linear combination of 6 radial orders of Zernike polynomials (Noll indices 4-36, so, $N_a=32$), following \eqref{eq: coronagraph entrance field}.
	When needed for the joint estimate, a $13\times8$ object source grid corresponding to the $N_p=104$ sky-angles for which the scene (exoplanet brightness coefficients) are to be estimated, is used.
	This source grid is a collection of point sources created such that their images propagated through the coronagraph are spaced by $1 \lambda / D$, and have an inner edge at $3 \lambda / D$ at the closest, and an outer edge at $10 \lambda / D$.
	The choice of this scene is to demonstrate how the techniques presented can recover both point sources (exoplanets) and extended sources (such as circumstellar disks), located at a few times the classical optics resolution limit of $\lambda/D$.
	The brightest point is $(1\times10^{-4})$ in the contrast units, while the faintest point is $(1.8\times10^{-6})$, with an average brightness of $3.9\times10^{-5}$.

	\begin{table}[htbp]
		\centering
		\begin{tabular}{| p{4cm} | p{1cm} | p{3cm} |}
			\hline
			\textbf{Simulation Parameter} & \textbf{Symbol} & \textbf{Value or Type} \\
			\hline
			Pixels in Entrance Pupil & $P$ & 1976 \\
			\hline
			True AO Residual & $\bw_t$ & vector of length $P$ \\
			\hline
			WFS Measurement & $\hat{\bw_t}$ &vector of length $P$ \\
			\hline
			Pupil Diameter & $D$ & 6.5 m \\
			\hline
			Coronagraph Operator & $\bD$ & Lyot \\
			\hline
			Fourier Transform Operator & $\mathcal{F}$ & \\
			\hline
			FPM & $\mathcal{M}$ & binary; \newline radius: $1.5 \lambda / D$ \\
			\hline
			Lyot Stop & $\mathcal{S}$ & binary; \newline diameter: $0.9D$  \\
			\hline
			WFS Model & $\mathcal{W}$ & Modulated PyWFS \\
			\hline
			PyWFS Modulation Radius & & $3 \, \lambda/D$ \\
			\hline
			Science Camera & SC & EMCCD \\
			\hline
			Pixels in SC & $L$ & 4489 \\
			\hline
			SC/WFS Cadence & & 1kHz, synchronized \\
			\hline
			SC Pixel Array Shape &  & $67 \times 67$ pixels \\
			\hline
			SC Pixels per $\lambda / D$ & & 0.39 \\
			\hline
			SC Readout Noise & $\sigma_{I_{ph}}$ &0.3 photon counts/px \\
			\hline
			Wavelength & $\lambda$ & Y band ($1 \mu m$) \\
			\hline
			Beam Splitter Ratio & & 70/30 \\
			\hline
			Source Apparent Magnitude &  & 8 (and 6) \\
			\hline
			Source Spectral Bandwidth & & 10\.\% \\
			\hline
			Photons per ms in SC & & $8,718$ ($55,008$) \\ 
			\hline
			Photons per ms in WFS & & $20,342$ ($128,351$) \\ 
			\hline
			Fried Parameter & $r_0$ & 0.3 m \\
			\hline
			Inner Scale & $l_0$ & 0.01 m \\
			\hline
			Outer Scale & $L_0$ & 25 m \\
			\hline
			Turbulence Model & & von K\'arm\'an; 6 layers; no scintillation; \newline  frozen flow \\
			\hline
			Strehl at $\lambda=1\mu m$ & & 0.73 \\
			\hline
			AO Correction Delay & & 2 Frames \\
			\hline
			NCPA Zernike Modes & & 4$-$36 (6 radial orders) \\
			\hline
			NCPA RMS & & 0.52 radian \\
			\hline
			True Object Grid & & $13 \times 8$ points \\
			\hline
			Grid Locations & & x: $[3 \lambda/D , 10 \lambda/D]$ \newline y: $[-6 \lambda/D , +6 \lambda/D]$ \newline Spacing of $1 \lambda/D$ \\
			\hline
			Relinearization Point & $x_{a,n}$ & iteration n linearization point \\
			\hline
		\end{tabular}
		\caption{\small Summary of the defining simulation parameters for the performed numerical experiments.}
		\label{Table: Simulation Parameters}
	\end{table}

	\section{Phase A}
	\label{sec:Compensating-NCPA}
	
	The objective of the first experiment performed is to estimate and compensate for the NCPA.
	In this experiment, we use \eqref{eq: y w/o planets} to estimate the coefficient vector $\ba$.
	Compensating for the NCPA using only one minute of observation time provides a good starting position for the more demanding combined estimation of the residual NCPA and the exoplanet image in "Phase B," described in Section~\ref{sec:EstimatingExoplanetImage}.
		
	\subsection{Estimating the NCPA}
	\label{ssec:EstimatingNCPA}

	\begin{figure}[htbp]
		\hbox{\includegraphics[width=.94\linewidth,clip=]{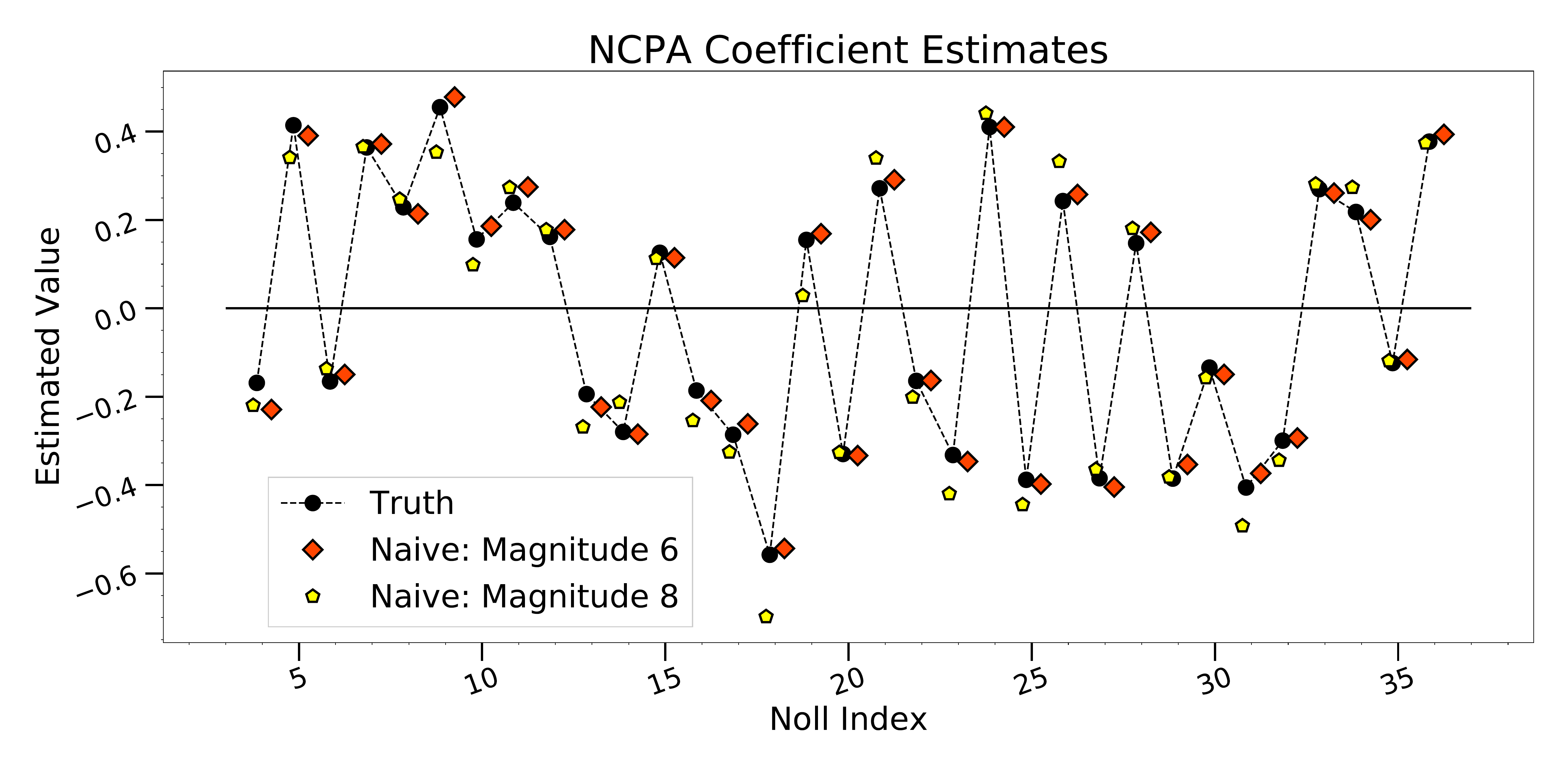}}
		\caption{\small The estimated aberration Coefficients from the simulated experiments, following five relinearization iterations. Included are the true, starting NCPA coefficients, and their na\"ive estimates using both a magnitude 6 (RMS error of $2.14\times10^{-2}$ radian) and magnitude 8 source (RMS error of $5.95\times10^{-2}$ radian).
		}
		\label{fig: EstimatedCoeffs}
		\vspace{-2mm}
	\end{figure}
	
	Phase A simulates a 1 minute observation, consisting of $60,000$ synchronized millisecond exposures in the science camera and WFS, the purpose of which is to provide a coarse estimate of large NCPA, which have an RMS phase of $0.52$ radian.
	The Phase A estimate uses the na\"ive regression model, which has unwanted bias that increases increases as the wavefront measurement error gets larger.
	The initial linearization point [see \eqref{eq: row gradient}] was taken to the zero vector, ie., $\ba_0 = 0$.
	The Phase A experiment was repeated for both a magnitude 6 and magnitude 8 target source.
	The AO system being simulated is not photon starved at magnitude 8, so although fewer photons are being observed compared to the magnitude 6 source, the Strehl ratio only decreases from 0.76 to 0.73 when going from the magnitude 6 source to the magnitude 8 source.
	The magnitude 8 source has a larger wavefront measurement error, which leads to larger bias in the na\"ive estimate.
	
	Fig.~\ref{fig: EstimatedCoeffs} shows the results of the na\"ive estimate of the aberration coefficients composing the NCPA, following a 5 iteration treatment of the nonlinearity (see Section~\ref{sec:Compensating-NCPA}.\ref{sssec:TreatingNonlinearity}), for both the magnitude 6 and 8 sources.
	The root mean squared (RMS) error of the magnitude 6 source estimates of the aberration coefficients is $0.0214$ radian, whereas for the magnitude 8 source the RMS error is $0.0595$ radian, approximately a factor of $3$ worse.
	
	\subsection{Treating the Nonlinearity}
	\label{sssec:TreatingNonlinearity}

	\begin{figure}[htbp]
		\hbox{\includegraphics[width=.94\linewidth,clip=]{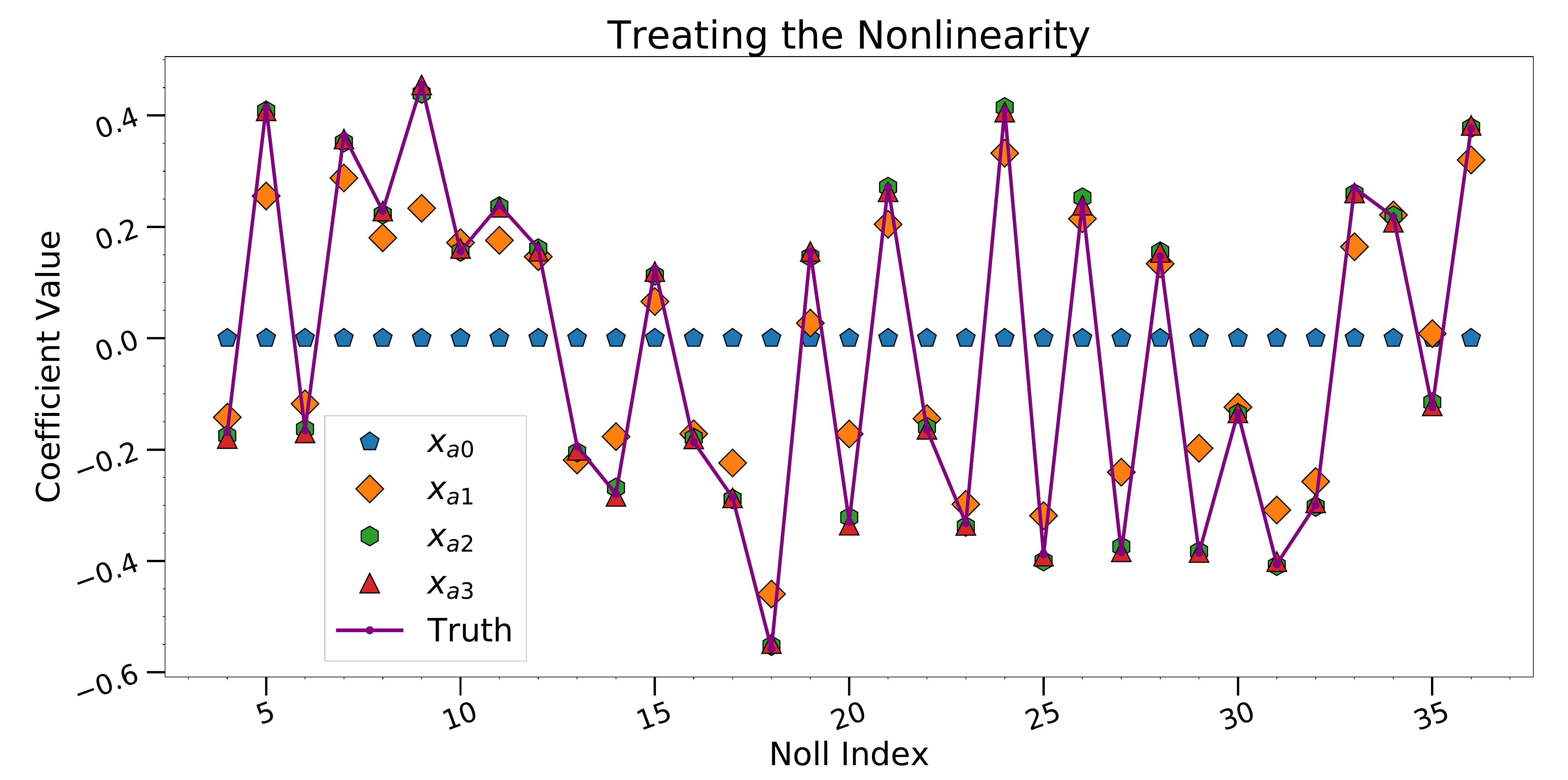}}
		\caption{\small Successive relinearization points, $x_{a,n}$, for the Phase A experiment, starting from all zeros, and then using the previous iteration's estimate.
		}
		\label{fig: RelinearizationSteps}
	\end{figure}

	In order to effectively treat the nonlinearity in the NCPA, the regression algorithm is run iteratively, successively relinearizing about the updated estimate values.
	Note that each iteration can be computed using the same set of observed data.
	The number of iterations that are needed can be determined by monitoring when the change in the updated linearization point from one iteration to the next becomes insignificant.
	This process is illustrated in Fig.~\ref{fig: RelinearizationSteps}, starting about the zero vector, and continuing using the previous iteration's estimate as the new linearization point.
	The only cost of adding relinearization steps is computation time since new observations are not needed.  
	In these simulations, three iterations effectively estimated the  aberration coefficients, but we applied five for good measure.

	\subsection{Compensating for the NCPA}
	\label{ssec:CompensatingNCPA}
	
	\begin{figure}[htbp]
		\centering
		\begin{subfigure}{}
			\centering
			\includegraphics[width=0.23\textwidth]{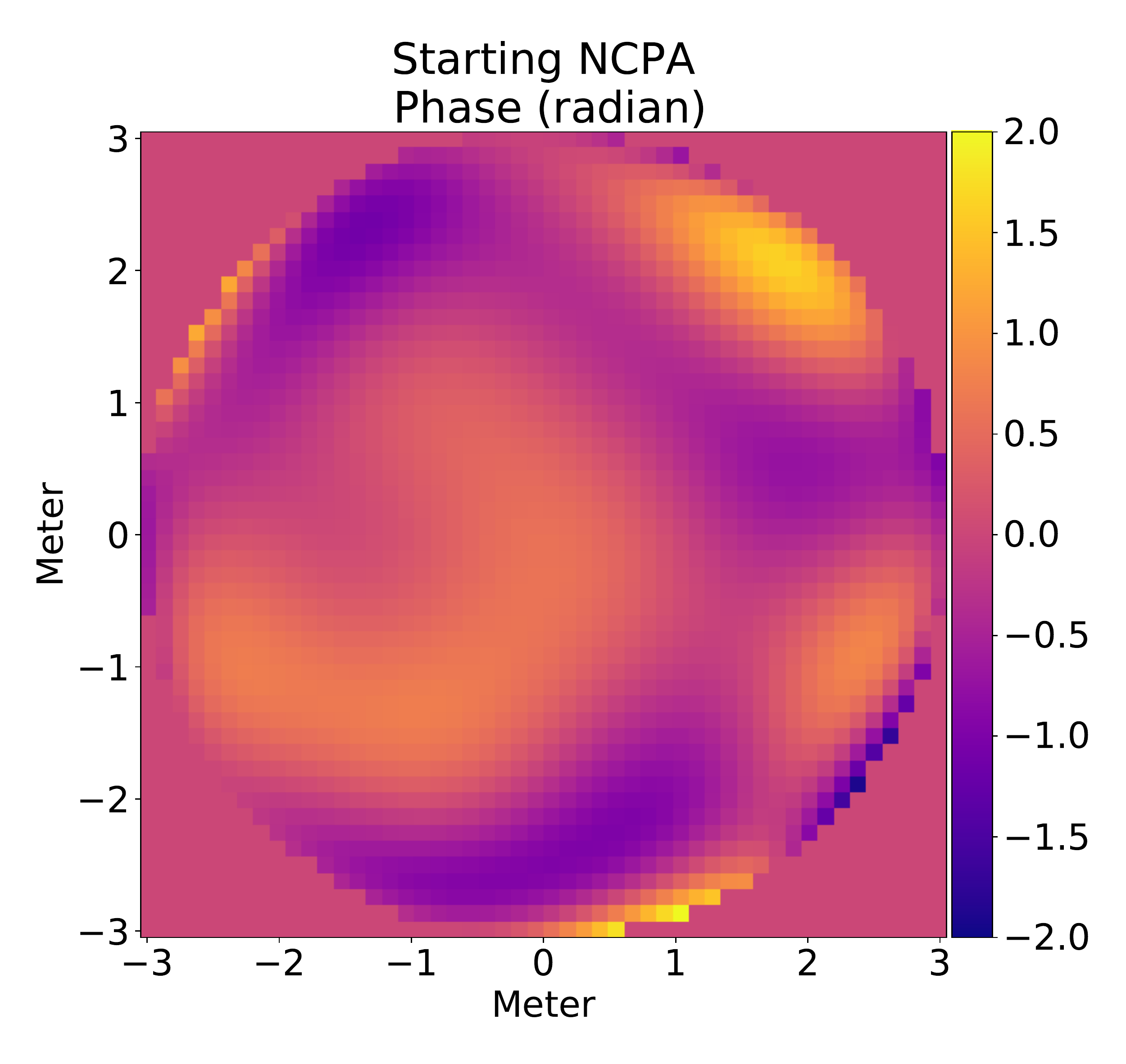}
		\end{subfigure}
		\begin{subfigure}{}
			\centering
			\includegraphics[width=0.23\textwidth]{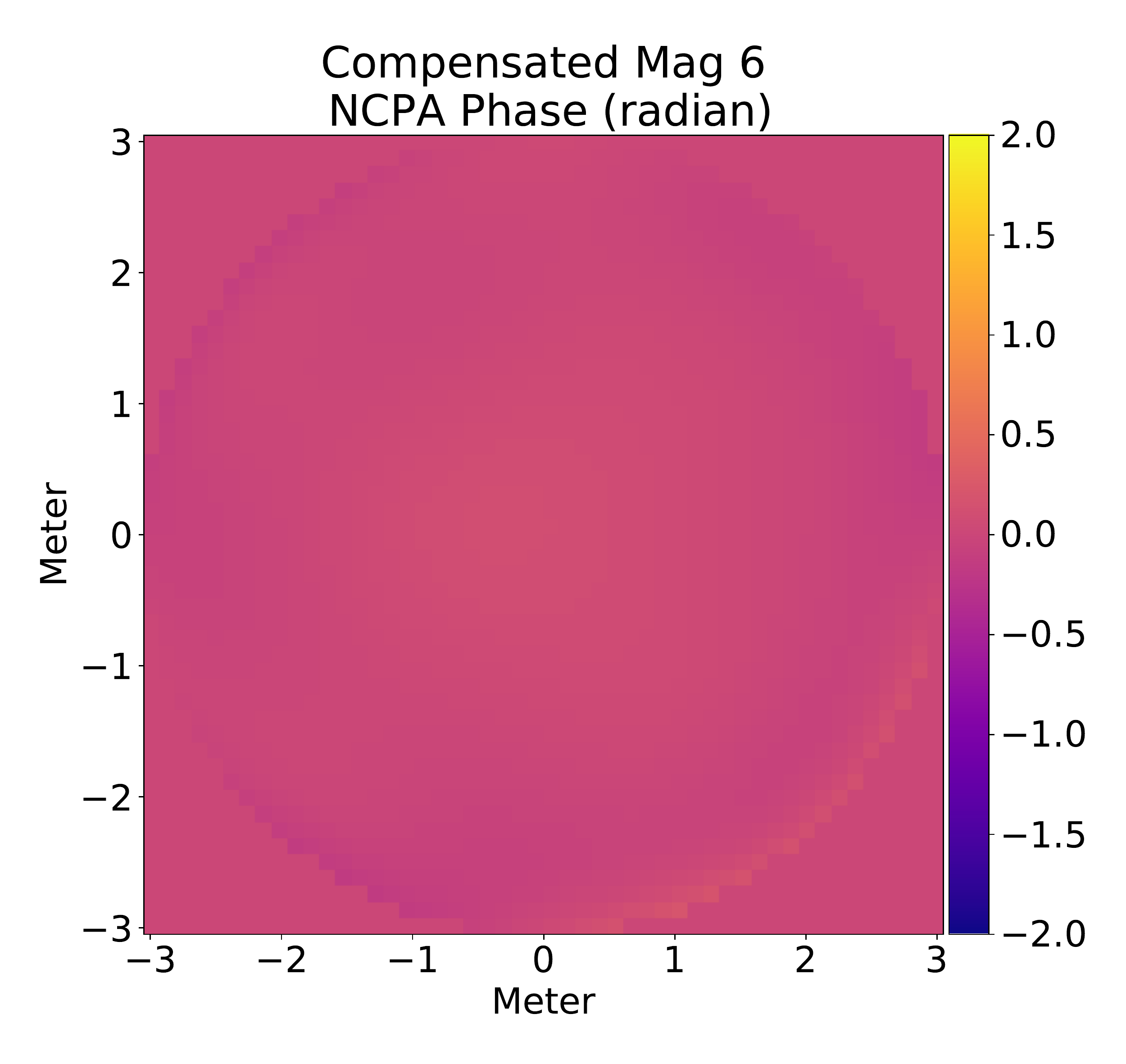}
		\end{subfigure}
		\begin{subfigure}{}
			\centering
			\includegraphics[width=0.23\textwidth]{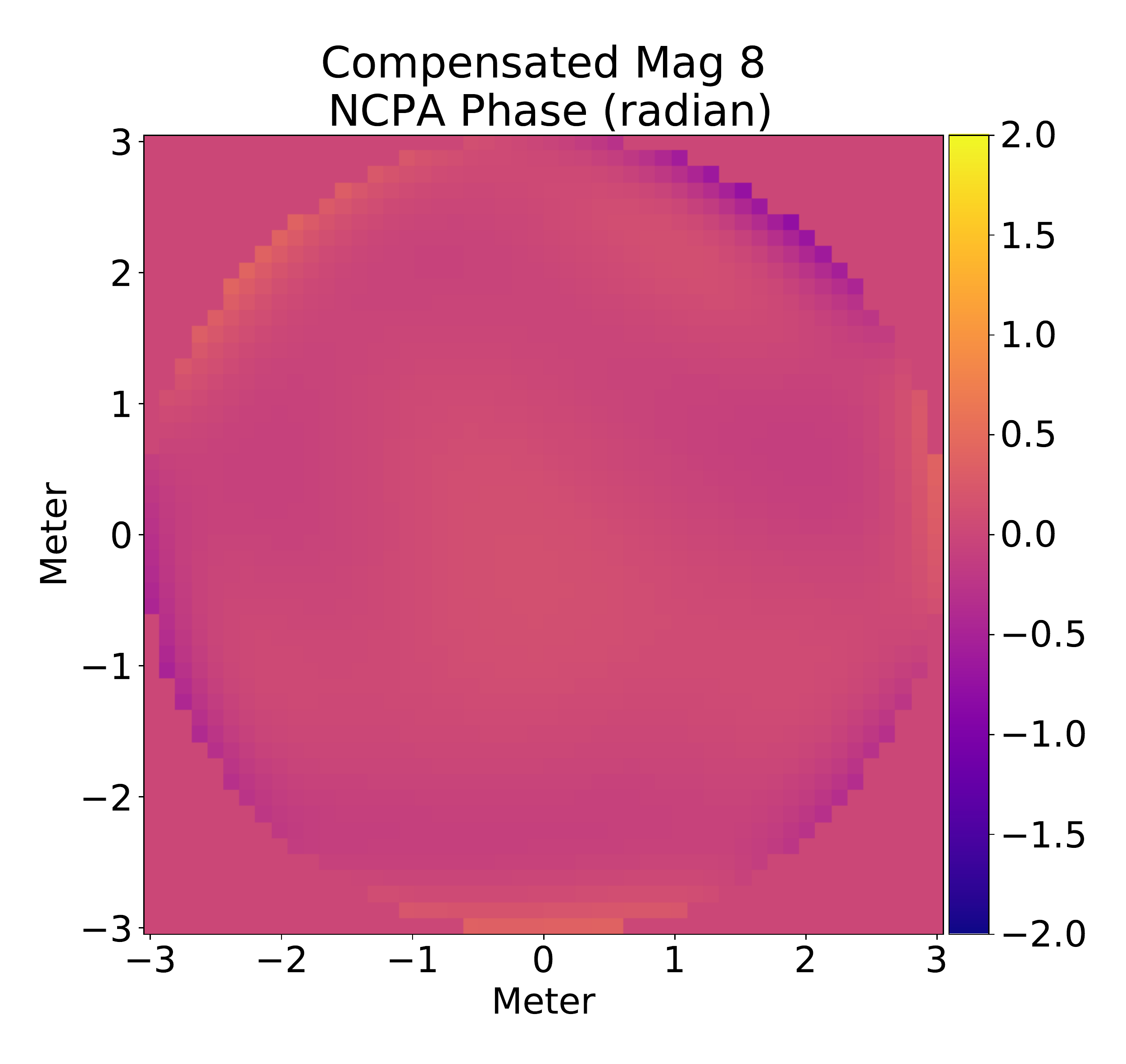}
		\end{subfigure}
		\caption{\small \emph{top left:} Starting NCPA phase (in radian), with RMS 0.52 radian, in the coronagraph entrance pupil. \emph{top right:} The phase in the coronagraph entrance pupil with the magnitude 6 source na\"ive estimate used for compensation. The residual phase has RMS 0.0495 radian. \emph{bottom:} The phase in the coronagraph entrance pupil with the magnitude 8 source na\"ive estimate used for compensation. The residual phase has RMS 0.11 radian.
		}
		\label{fig: Phase1_NCPA}
		\vspace{-2mm}
	\end{figure}
	
\begin{figure}[htbp]
		\hbox{\includegraphics[width=.94\linewidth,clip=]{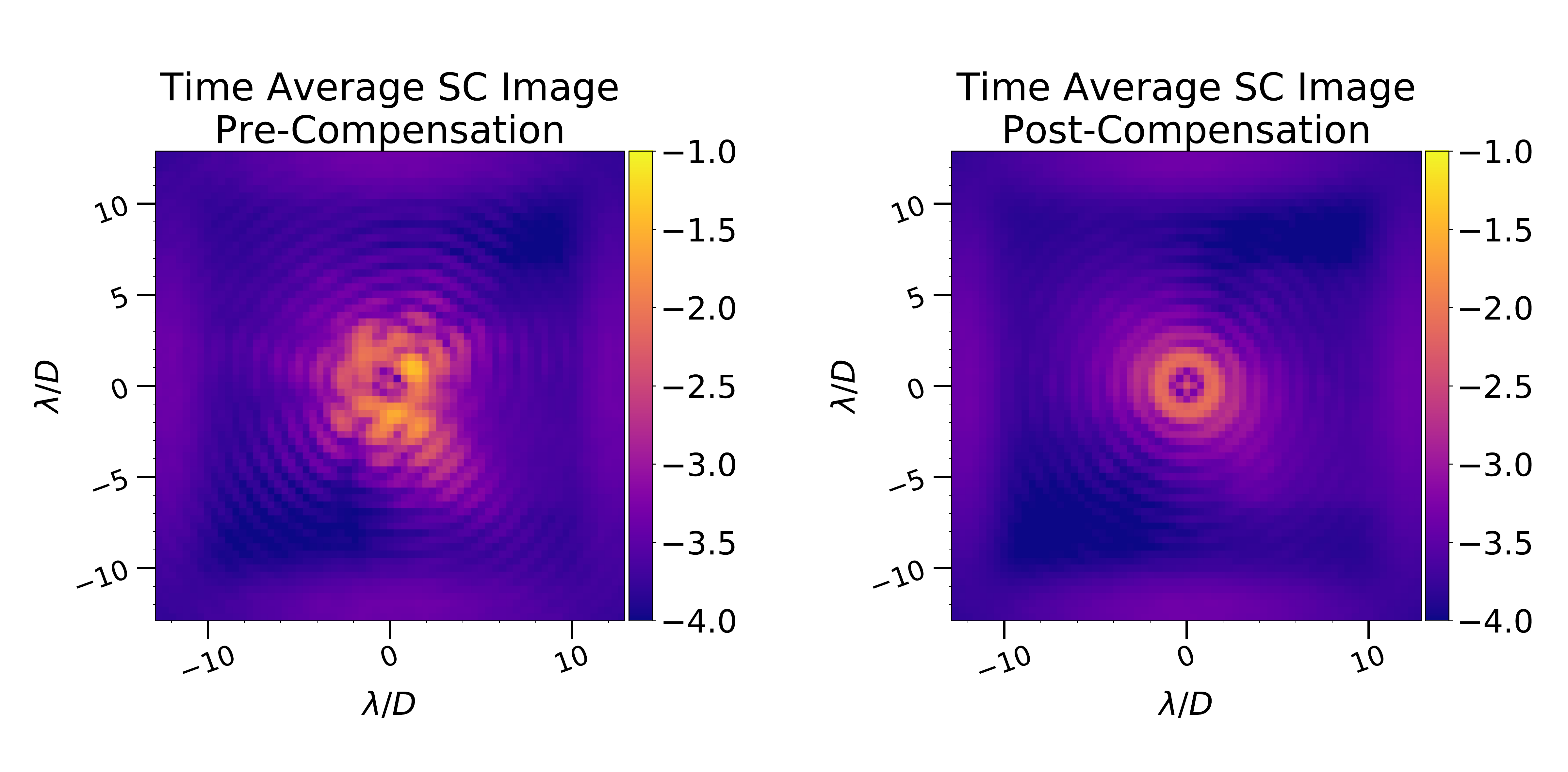}}
		\caption{\small Log scale Time averaged images in the science camera for the Phase A and Phase B experiments in contrast. \emph{left:} the manifestation of the $0.52$ radian RMS ($\boldsymbol{\approx \lambda / 13}$) NCPA is present, and shows a significant degradation to the coronagraphic image. \emph{right:} the estimated NCPA has been applied to a DM to compensate, leaving behind the manifestation of a $0.0495$ radian RMS residual NCPA.
		}
		\label{fig: Phase1_SC}
		\vspace{-2mm}
	\end{figure}

	Once we have estimated the NCPA, we can use a non-common path DM (such as the one in MagAO-X [\citenum{Close2018_MagAOXdesign}]) to compensate for them. 
	A DM command that performs the  compensation can be calculated by a least-squares fit of the NCPA optical path difference (OPD) map to the DM command modes.
	The resulting phase maps in the coronagraph entrance pupil after DM compensation for our simulations are shown in Fig.~\ref{fig: Phase1_NCPA} in the top right (magnitude 6 source) and bottom (magnitude 8 source) panels.
	As we expect from the the RMS error of the estimated aberration coefficients, compensating the NCPA via the magnitude 6 source estimate does a better job, achieving a reduction in RMS phase due to the NCPA from $0.52$ radian to $0.0495$ radian, a $10.5\times$ improvement.
	Using the magnitude 8 source estimate achieves a reduction in RMS phase from $0.52$ radian to $0.11$ radian, a $4.74\times$ improvement.
	In order to proceed in Phase B, we will adopt the better compensated NCPA while the magnitude 8 source in order to demonstrate being able to estimate a small NCPA jointly with the exoplanet image.
	This choice could represent an observation sequence where the NCPA is calibrated on a bright star, and then the telescope would slew to a fainter science target.
	If this is done, any change to the NCPA due to the slewing would likely remain small, especially if the calibration star is close to the target on sky.
	If additional NCPA is incurred through this slewing, it can be estimated while observing the target star, as will be discussed in the following section.
	The resulting time averaged images in the science camera in contrast units for the cases of pre- and post compensation, can be seen in Fig.~\ref{fig: Phase1_SC}.

	\begin{table}[htbp]
		\centering
		\begin{tabular}{| p{2cm} | p{2cm} | p{2.85cm} |}
			\hline
			\textbf{Magnitude of Source} & \textbf{RMS Error (radian)} & \textbf{Compenasted RMS Phase (radian)} \\
			\hline
			\centering
			6 & $0.0214$ & $0.0495$\\
			\hline
			\centering
			8 & $0.0595$ & $0.11$ \\
			\hline
		\end{tabular}
		\caption{\small Summary of the resulting root mean square (RMS) error using the na\"ive estimate based on 1 minute of observations to solve for the NCPA, and the residual RMS phase left after using the estimates to compensate the NCPA. Note the starting RMS phase was $0.52$ radian.}
		\label{Table: Phase A Experiment Results}
	\end{table}
	
	\section{Phase B: Joint Estimation}
	\label{sec:EstimatingExoplanetImage}
	
	The second experiment performed is to jointly monitor the residual NCPA left over after Phase A, while also estimating any present exoplanetary image.
	Recalling the results from Section\ \ref{sec:Compensating-NCPA}, the NCPA present in the coronagraph optical train has an RMS phase error of $0.0495$ radian, and the time average image in the science camera can be seen in the right frame of Fig.~\ref{fig: Phase1_SC}.
	Phase B is a $4$ minute observation of a magnitude 8 star taking place directly after the $1$ minute Phase A observation described in Section\ \ref{sec:Compensating-NCPA}.
	Phase B gathers $240,000$ millisecond synchronized exposures in both the WFS and science camera telemetry streams.
	The same process of treating the nonlinearity described above is performed here, starting by linearizing the regression equations with the $N_a$ NCPA coefficients and $N_p$ exoplanet coefficients set to zero (see \eqref{eq: y}), and successively updating with the previous iteration bias-corrected estimate, for a total of five iterations.
	Note that after compensating for the NCPA with the DM, the average level of stellar light in the science camera where the exoplanet image is being estimated is $\approx 3\times10^{-4}$ in contrast units.
	
	\subsection{The Na\"ive Estimate}
	\label{ssec:NaiveEstimate}
	
	\begin{figure}[htbp]
		\hbox{\includegraphics[width=.94\linewidth,clip=]{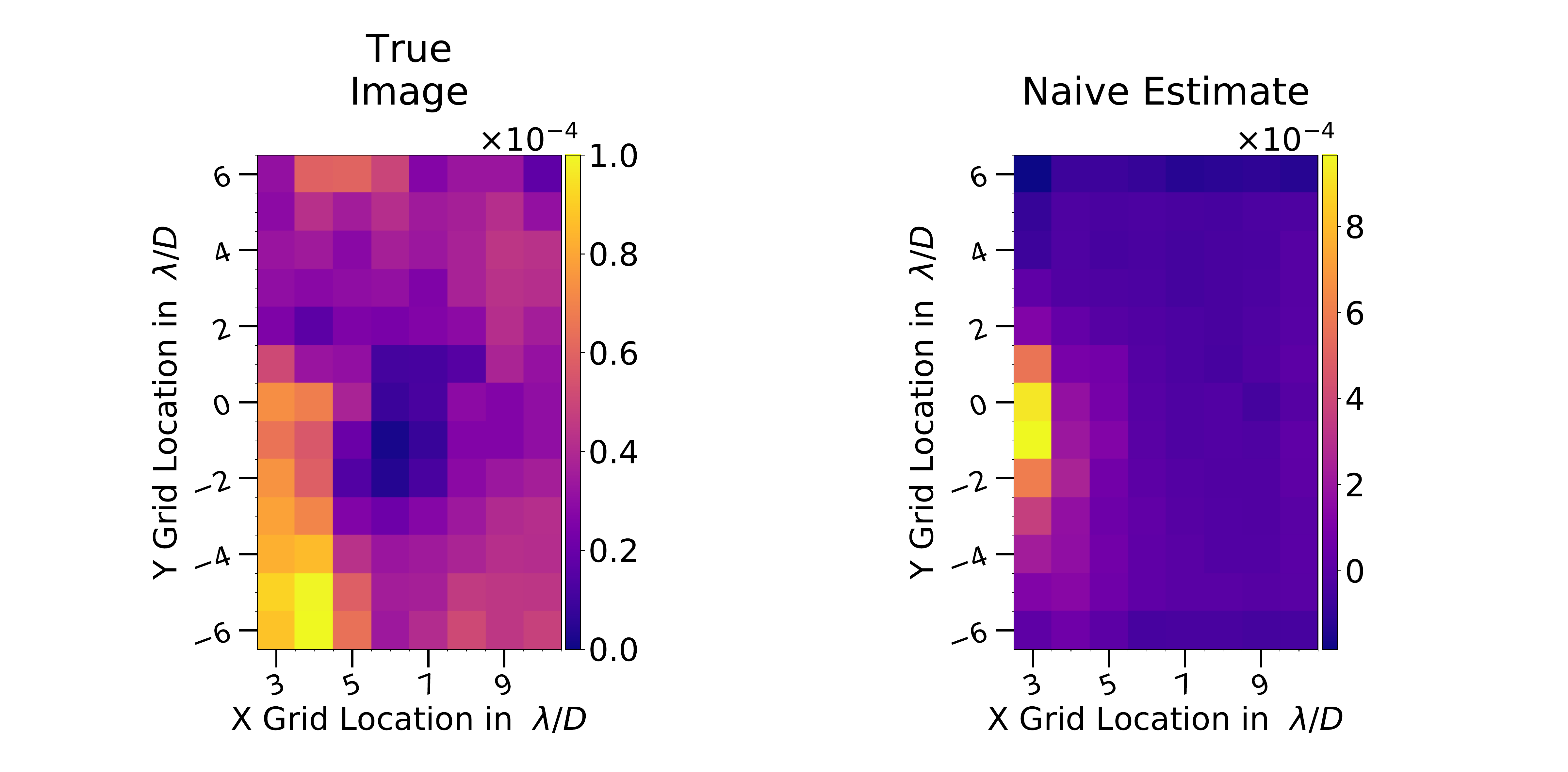}}
		\caption{\small The estimated exoplanet imaging coefficients from the simulated Phase B experiment, following five relinearization iterations. \emph{left:} the true object source grid. \emph{right:} Na\"ive estimate of object source grid. All units are in contrast.
		}
		\label{fig: NaiveEstImages}
		\vspace{-2mm}
	\end{figure}
	
	As we have seen in the Phase A experiment, the na\"ive estimate can perform quite well.
	A combination of factors including but not limited to WFS measurements of reasonable quality and the large NCPA, provided conditions under which the na\"ive estimate recovers a useful estimate of the NCPA.
	However, we found that when using the na\"ive estimate to solve for the exoplanet image coefficients and smaller NCPA coefficients, the bias from the WME renders the estimate unacceptable, as can be seen in Fig.\ref{fig: NaiveEstImages} from the Phase B experiment.
	In this figure, the estimated imaging coefficients have been reshaped and plotted as a $13\times8$ image to allow for easy comparison to the object source grid.
	It is clear this is not a useful estimate of the exoplanet image, as it results in an RMS error in brightness (contrast units) of $1.63\times10^{-4}$, not to mention the fact that the morphology of the image is almost entirely lost.

	\subsection{The Bias-Corrected Estimate}
	\label{ssec:CSNEstimate}
	
	\subsubsection{Choosing Monte Carlo Wavefronts}
	\label{sssec:MCSimpleMethod}
	As described in Part II, the bias-corrected estimator uses a set of Monte Carlo wavefronts created using the knowledge of the spatial statistics of the AO residual wavefronts.
	To generate the Monte Carlo wavefronts, we calculate the mean and covariance matrix of all of the $T$ AO residual wavefront vectors, the set of which is represented as $\{ \bw_t \}$.
	While it is not possible to know the exact mean and covariance matrix of the AO residuals on a real system (since only the measurements $\{ \hat{\bw}_t \}$ are available), we ignore that complication for the purposes of this study.
	We then draw the Monte Carlo wavefronts from a $P$-dimensional multivariate normal with sample mean and covariance of the AO residual wavefronts.
	The Monte Carlo wavefronts are thus not temporally correlated like the true AO residual wavefronts, but there is no need for them to be, so long as their distribution has the same moments  as the time-averaged moments of the stochastic process governing the true AO residual wavefronts.
	
	Although our Monte Carlo wavefronts have the same 2\underline{nd} order statistics as the $T=240,000$ AO residual wavefronts in the $4$ minute on-sky data set, (apart from the error due to the finite number of Monte Carlo samples, which is mitigated by using $480,000$ separate Monte Carlo samples), the accuracy of the bias-corrected estimate is limited by the fact that the true AO residual wavefronts did not obey multivariate normal statistics.
	We know this because the univariate (i.e., single pixel) AO residual values failed the Anderson-Darling test for univariate normality.
	(Univariate normality of all of the individual variables is a necessary, but not sufficient, condition for multivariate normality.)
	These statistical details are discussed in Part II.

	\subsubsection{Estimating the Exoplanet Image}
	\label{sssec:Imaging}

		\begin{figure}[htbp]
		\hbox{\includegraphics[width=.94\linewidth,clip=]{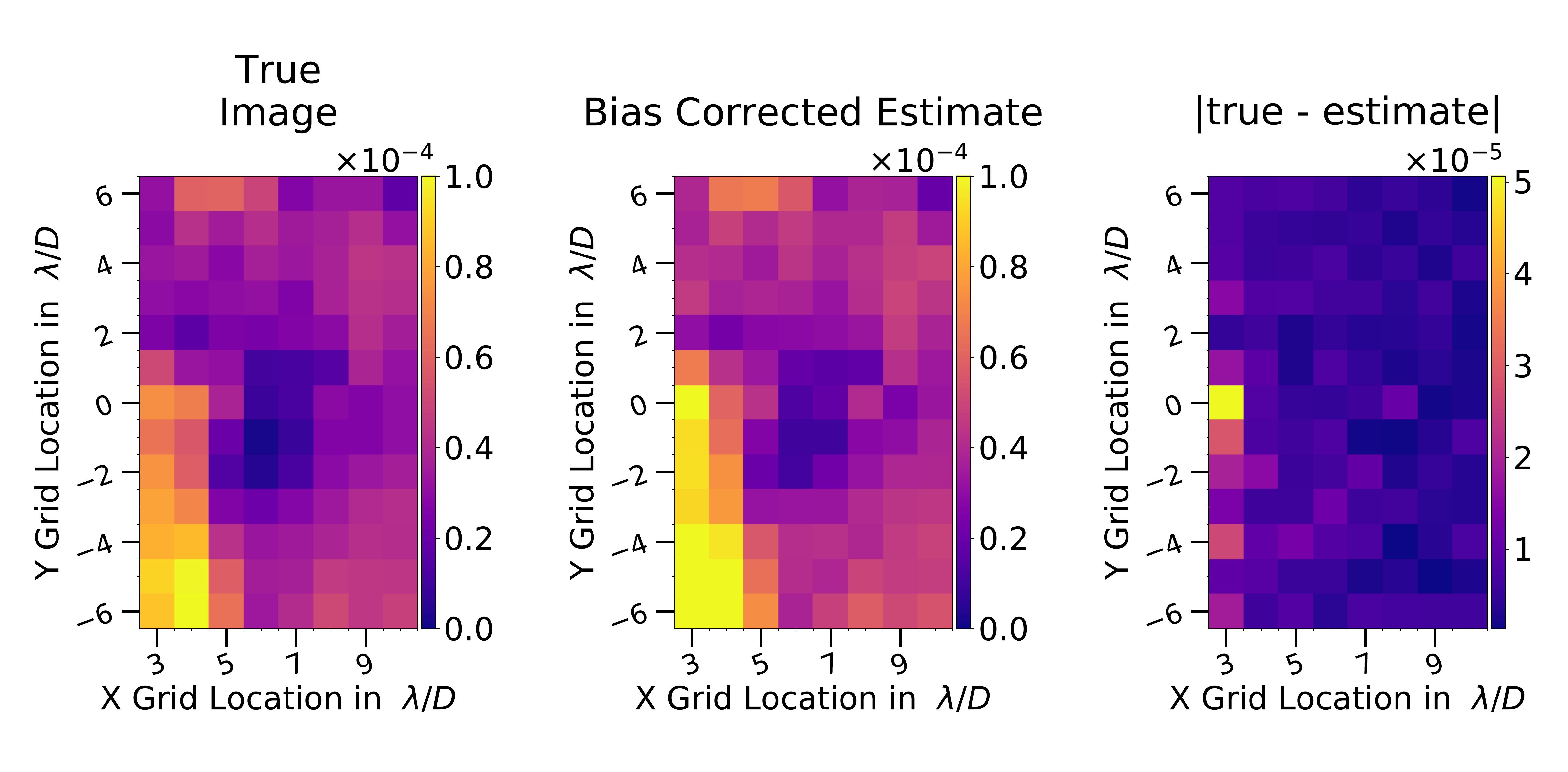}}
		\caption{\small The estimated exoplanet imaging coefficients from the simulated Phase B experiment, following five relinearization iterations, reshaped in to the $13\times8$ grid. Left: the true object source grid. Middle: Bias-corrected estimate of object source grid. Right: The absolute value of the error in the estimate. All units are in contrast.
		}
		\label{fig: CSNEstImages}
		\vspace{-2mm}
	\end{figure}

	Performing the bias-corrected estimate shows a marked improvement over the na\"ive estimate, recovering much of the detail in the true signal, as can be seen in Fig.~\ref{fig: CSNEstImages}.
	The true image is in the left frame, the bias-corrected estimate is in the middle, and the error in the estimate is in the right frame.
	The RMS error for the bias-corrected estimate of the exoplanet brightness coefficients, averaged over all of the points is $9.5\times10^{-6}$, a $17.25\times$ improvement over the na\"ive estimate.
	Looking more closely at the error in the estimated coefficients, there are three grid points at a distance near $3 \, \lambda/D$, where the PSF is bright, with an RMS error of about $3.5 \times 10^{-5}$.  The 13 grid points with a distance near $10 \, \lambda/D$, where the PSF is not a bright, have an RMS error of about $4.2 \times 10^{-6}$.  Note these the RMS errors in the estimates are indicative of the $1 \, \sigma$ contrast achieved, since the estimate error has very little dependence on the exoplanet brightness (assuming that it is much less bright than the star). 
	The ideal estimates of the exoplanet coefficients are better by roughly a factor of 20.
	Note that we did not attempt to reduce the PSF brightness with dark hole techniques (in principle, dark hole techniques should be able to leverage the NCPA estimates).
	

	\begin{figure}[htbp]
		\centering
		\begin{subfigure}
			\centering
			\hbox{\includegraphics[width=.94\linewidth,clip=]{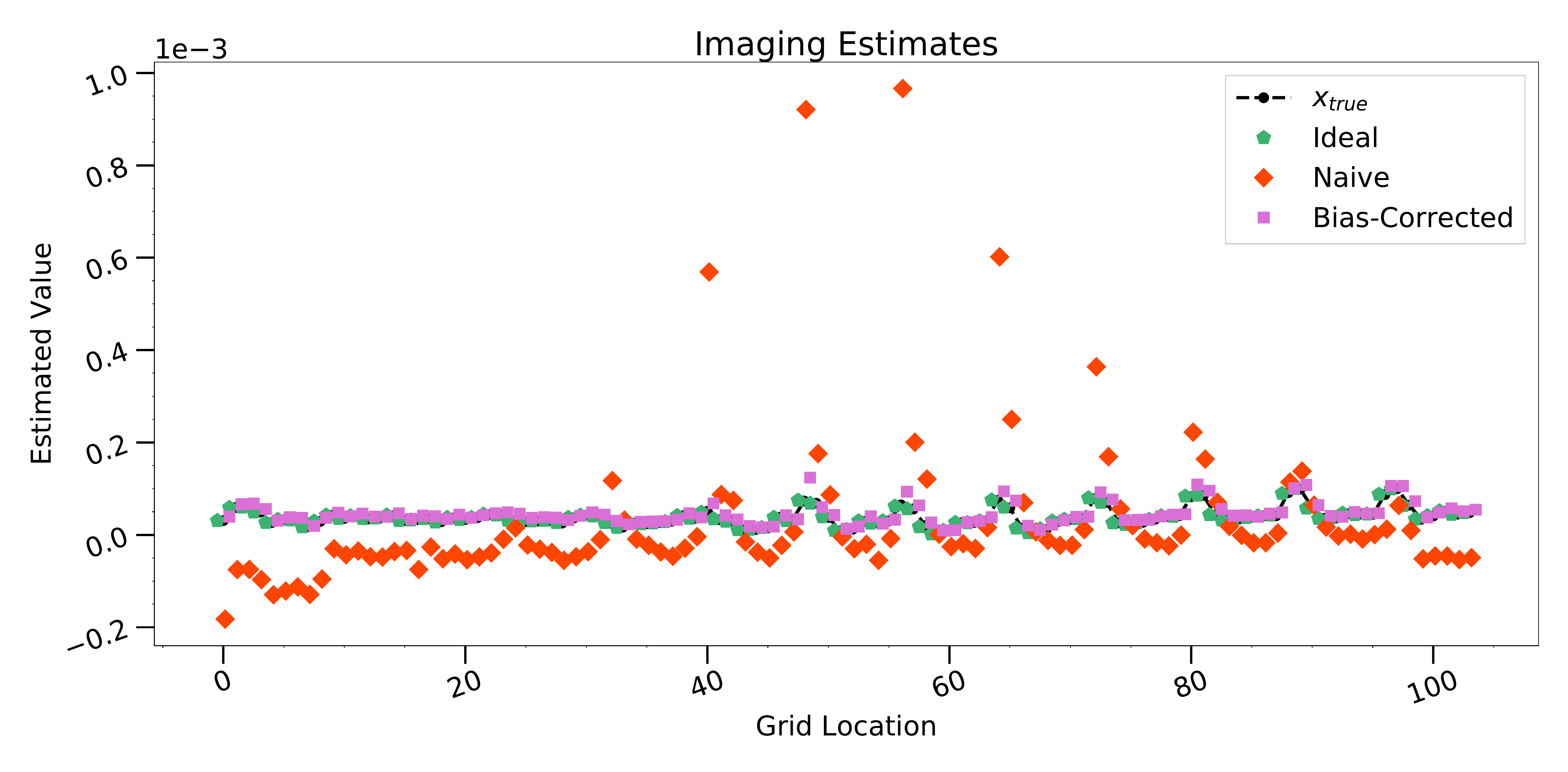}}
		\end{subfigure}
		\begin{subfigure}
			\centering
			\hbox{\includegraphics[width=.94\linewidth,clip=]{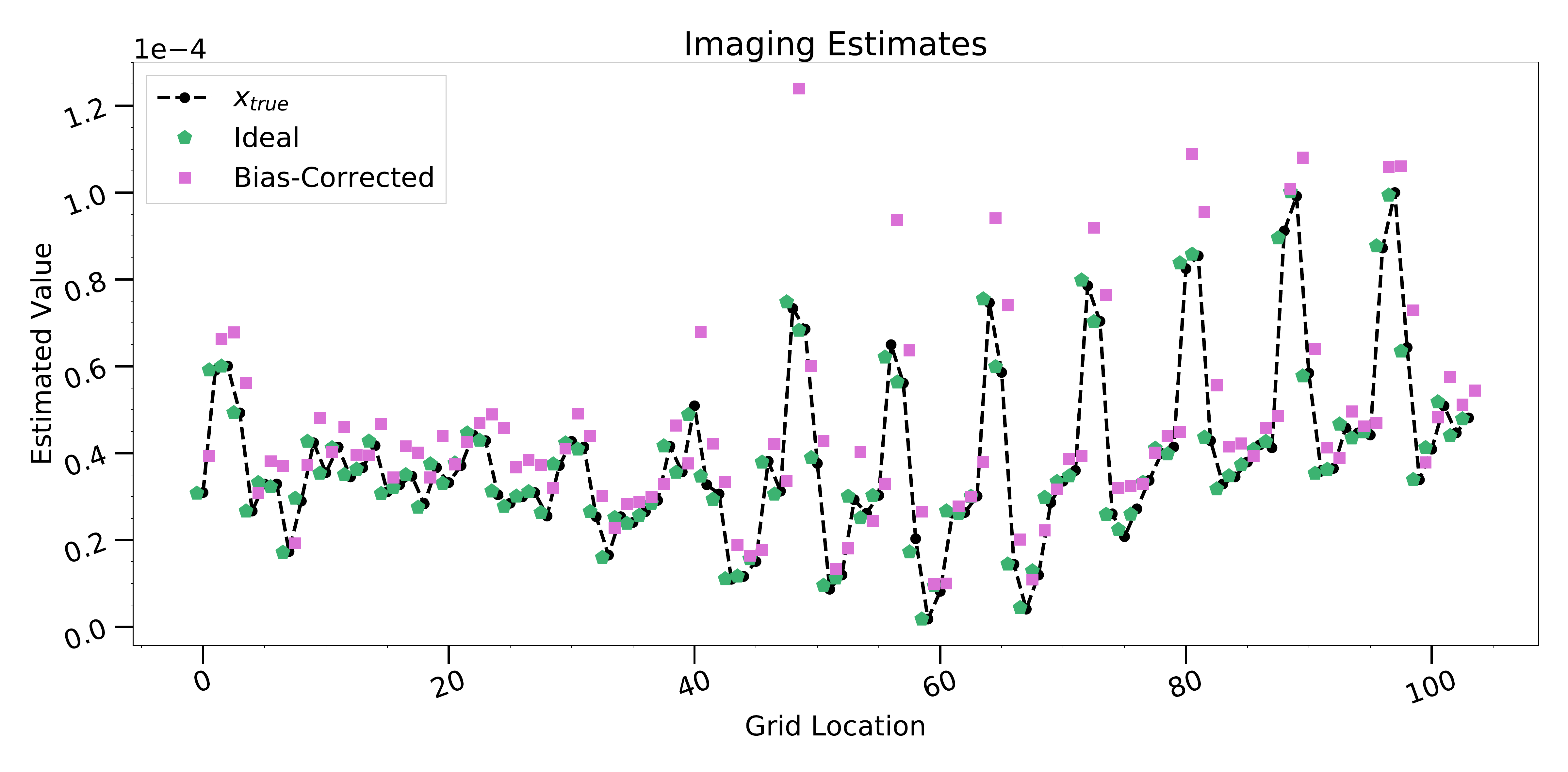}}
		\end{subfigure}
		\begin{subfigure}
			\centering
			\hbox{\includegraphics[width=.94\linewidth,clip=]{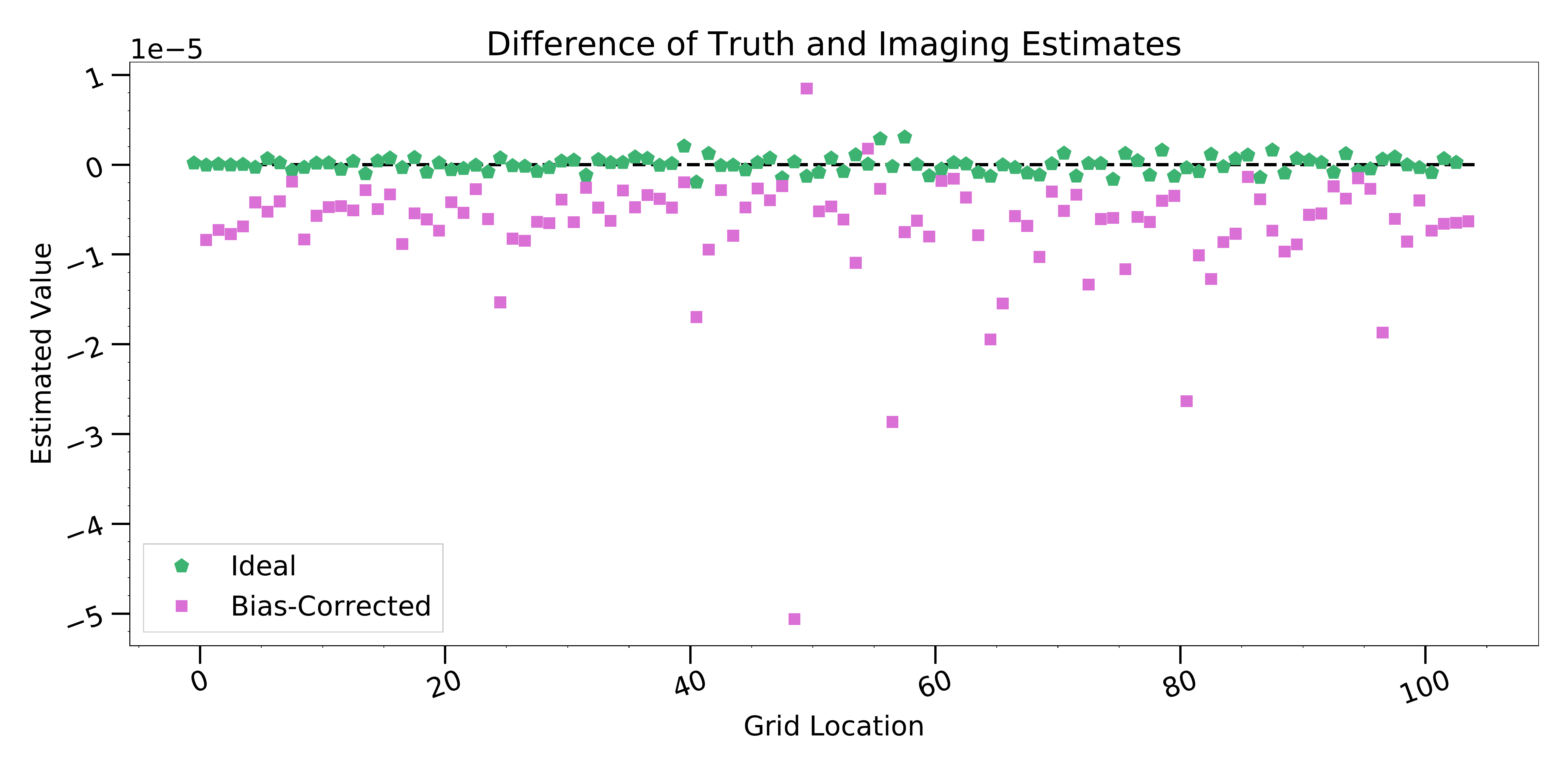}}
		\end{subfigure}
		\caption{\small \emph{top:} The exoplanet brightness coefficients for each of the three estimators for the Phase B experiment, following five relinearization iterations. \emph{middle:} The same as the top, except the na\"ive estimates are not displayed. The error bars provided by the ECM are too small to be seen. The partial $\chi^2$ value for the ideal estimate is 0.76, and for the bias-corrected estimate it is 225. \emph{bottom:} The difference of the estimated coefficients and the true values for both the ideal and bias-corrected estimators}
		
		\label{fig:EstImagingCoeffs}
		\vspace{-2mm}
	\end{figure}
	
	The vector of exoplanet image coefficients is plotted in Fig.~\ref{fig:EstImagingCoeffs}.
	The top frame shows the coefficients for each of the estimates as well as the truth values, and the bias of the na\"ive estimate is rather evident.
	In order to see the much smaller residual bias in the bias-corrected estimate, the middle frame in this figure does not contain the na\"ive estimate values.
	The bottom frame in this figure shows the estimated values subtracted from the true values.
	As explained above, this residual bias in the bias-corrected estimate is largely due to the fact that the true AO residuals do not obey multivariate normal statistics, while the Monte Carlo wavefronts do.
	
	In principle (i.e., assuming that the Monte Carlo wavefronts have the same statistics as the true AO residual wavefronts), the bias-corrected estimate should be nearly as accurate as the ideal estimate, but here we see that the ideal estimate (which is unbiased) is very close to the true values, with an RMS error in contrast units of $8.83\times10^{-7}$.
	The plots in Fig.~\ref{fig: CSNEstImages} and corresponding RMS error numbers show that the simple assumption that the statistics of $\bw$ are governed by a multivariate normal, with mean and covariance matrix calculated from the $4$ minute set of millisecond wavefronts, is a useful assumption (at least when compared to na\"ive estimation), but that it is not adequate to remove all of the bias.

	\subsubsection{Comparing to Perfect PSF Subtraction}
	\label{ssec:PSFsubtraction}
	
	\begin{figure}
		\hbox{\includegraphics[width=.94\linewidth,clip=]{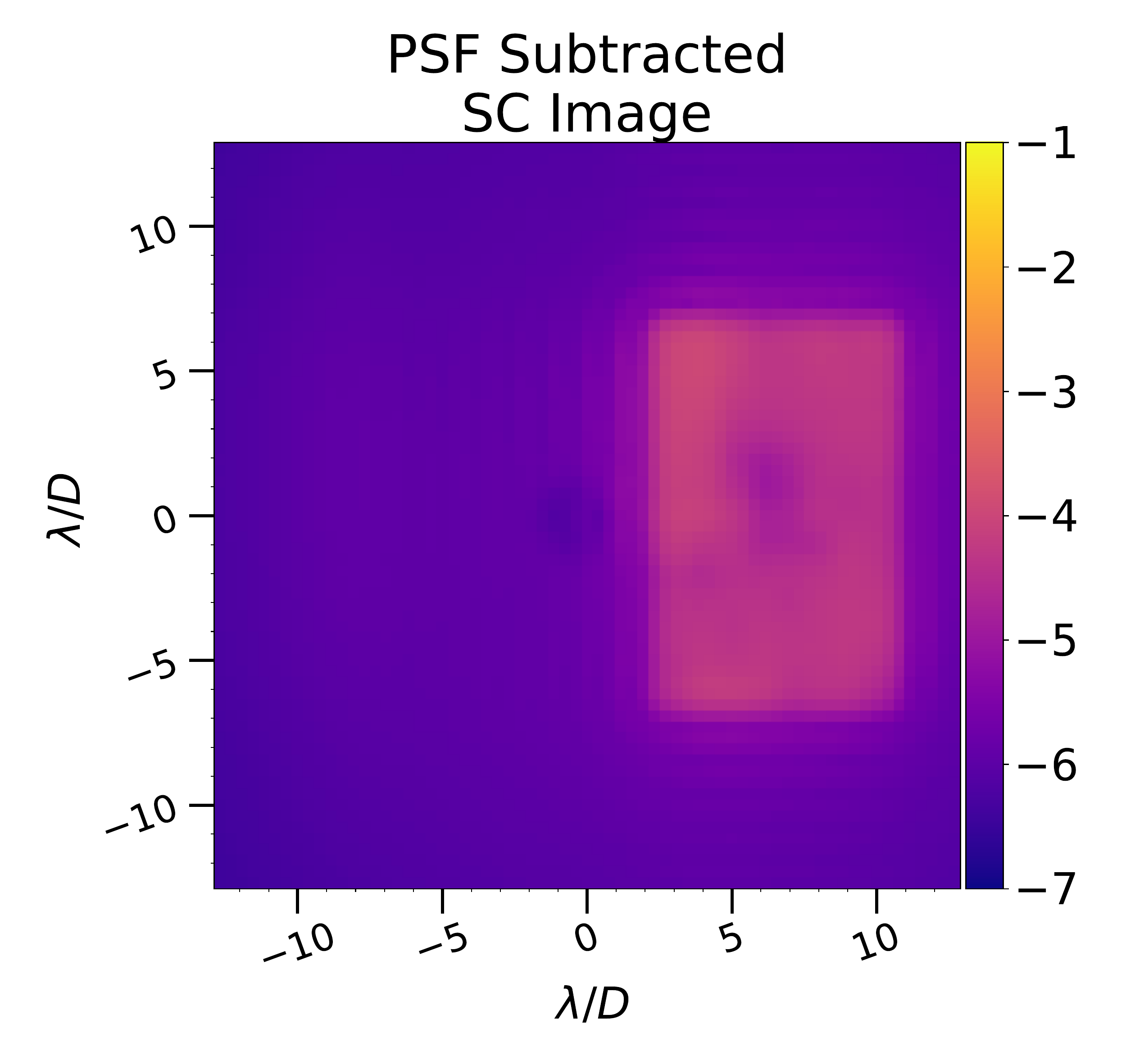}}
		\caption{\small Log scale focal plane image in contrast units after subtracting a perfect PSF from the science image. The perfect PSF is created by using the same sequence of wavefronts as the science image.
}
		\label{fig: Phase2PSFsubSC}
		\vspace{-2mm}
	\end{figure}

	In this section, we compare the bias-corrected estimate with a more common method, PSF subtraction.
	To proceed with this discussion, we first define some terms that will be used.
	PSF stands for "point-spread function," which for our purposes amounts to the time-average image in the science camera that would be observed without any planets.  It carries the effects of diffraction, NCPA and the AO residual wavefronts.
	The \emph{science image} is the time-average of the science camera telemetry, and it is shown in the right-hand frame of Fig.~\ref{fig: Phase1_SC}.
	We further define the \emph{perfect PSF}:
\begin{itemize}
	\item it is created using the same sequence of wavefronts as the science image
	\item it has the exact same NCPA as the science image
	\item it has the same noise realizations as the science image
	\item it excludes all light coming from the circumstellar material (``the exoplanet scene”).
\end{itemize}
	Thus, the perfect PSF is the ultimate image to subtract from the science image, as it perfectly removes starlight and noise.
	In particular, subtracting the perfect PSF does not suffer from self-subtraction artifacts associated with differential imaging (see Section~\ref{sec:currentTechniques}.\ref{ssec:DI}), and which arise from having to measure a PSF to subtract on-sky.
	Of course, the perfect PSF cannot be known outside of simulation studies, but it is a useful benchmark.
	Fig.~\ref{fig: Phase2PSFsubSC} shows the resulting difference image of subtracting the perfect PSF from the science image following the convention described in Section\ \ref{sec:NumericalModels}.\ref{ssec:Coronagraph}.

\begin{figure}
		\hbox{\includegraphics[width=.94\linewidth,clip=]{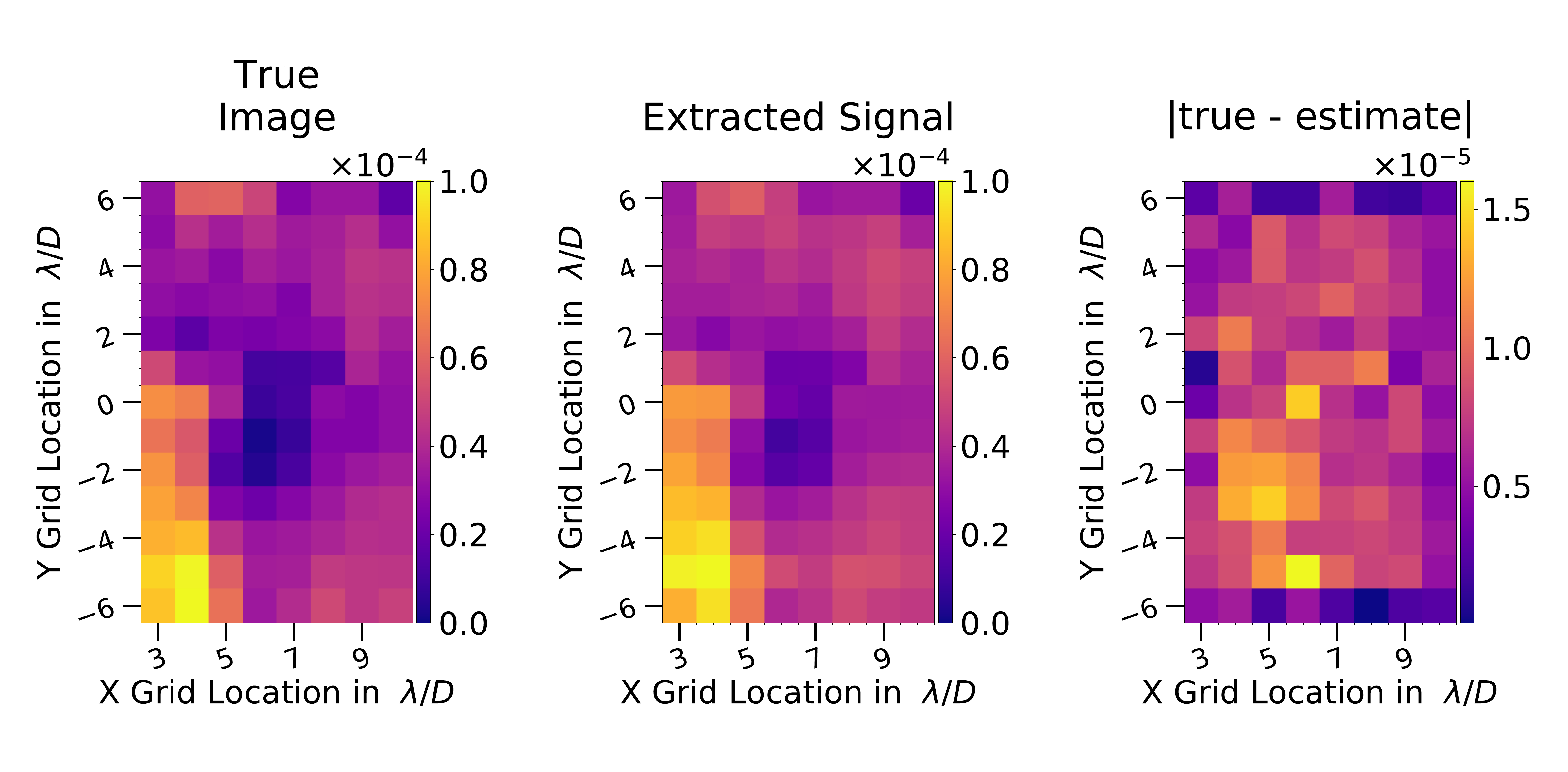}}
		\caption{\small Extracting the exoplanet signal from the PSF subtracted focal plane. Left: The true object source grid. Middle: Extracted signal from doing perfect PSF subtraction. Right: The absolute value of the error in the extracted signal. All units are in contrast.
		}
		\label{fig: Phase2PSFsubimages}
		\vspace{-2mm}
	\end{figure}

Despite the perfect PSF subtraction, the image in Fig.~\ref{fig: Phase2PSFsubSC} is missing much of the detail present in true image seen in left frame of Fig.~\ref{fig: CSNEstImages}.
This is due to the fact that the image is blurred by the off-axis PSF.
While one could attempt deconvolution to mitigate the blurring, instead, we used our knowledge from the simulation that the exoplanet image consists of point sources located on a grid.
This allowed us to sidestep deconvolution simply by extracting the values of the image in Fig.~\ref{fig: Phase2PSFsubSC} at the grid locations.
The result of that extraction is seen in the middle frame of Fig.~\ref{fig: Phase2PSFsubimages}, with the true source grid pictured for convenience in the left frame, and the corresponding error in the right frame.
	The subsequent measurement of the RMS error for this extraction is $7.63\times10^{-6}$.
	Referring to Table\ (\ref{Table: Phase B Experiment Results}), this result can be directly compared to the RMS error for the various estimate techniques presented, showing that the perfect PSF subtraction result quite is comparable to that of the bias-corrected estimate, with the RMS error of the bias corrected estimate being about 25\% larger.
	The RMS error of the ideal estimate turned out to be $8.6\times$ better than the perfect PSF subtraction.
	
	While we were able to use knowledge of point-source locations to sidestep the deconvolution issue after performing the PSF subtraction, this would not be possible on-sky, and it is important to understand the our estimation methods automatically include the blurring effects that off-axis PSF has on the planetary image.
	

	\begin{table}[htbp]
	\centering
	\begin{tabular}{| p{2.25cm} | p{2.25cm} | p{2.75cm} |}
		\hline
		\textbf{Estimate Type} & \textbf{NCPA Estimate RMS error \newline (radian)} & \textbf{Exoplanet Image Estimate RMS \newline error (brightness)} \\
		\hline
		Na\"ive & $1.13\times10^{-2}$ & $1.63\times10^{-4}$\\
		\hline
		Bias-Corrected & $4.39\times10^{-3}$ & $9.45\times10^{-6}$ \\
		\hline
		Ideal & $1.82\times10^{-4}$ & $8.83\times10^{-7}$ \\
		\hline
		Perfect PSF \newline subtraction & \centering N/A & $7.63\times10^{-6}$ \\
		\hline
	\end{tabular}
	\caption{\small Error metrics for a simulated $4$ minute observation.  Summary of the resulting root mean square (RMS) error using the na\"ive, bias-corrected, and ideal estimates to jointly solve for the NCPA and exoplanet image. The results for using ideal PSF subtraction for estimating the exoplanet image are also included}
	
	\label{Table: Phase B Experiment Results}
\end{table}

	\subsection{Jointly Monitoring the NCPA}
	\label{ssec:Imaging-JointNCPA}
	
	\begin{figure}[htbp]
		\hbox{\includegraphics[width=.94\linewidth,clip=]{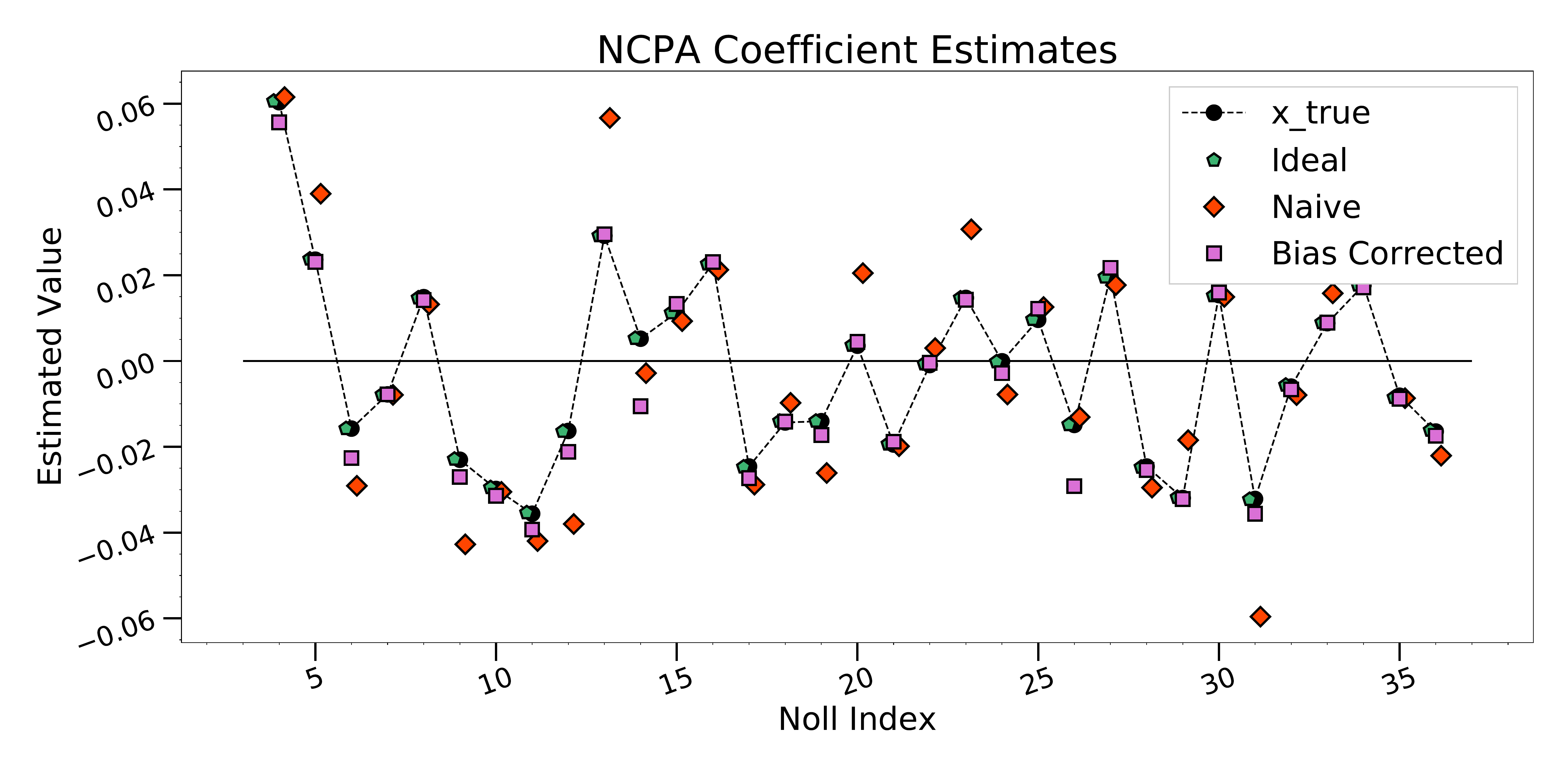}}
		\caption{\small The joint estimates of the NCPA coefficients found while estimating the image in the $4$ minute Phase B experiment. The error bars provided by the ECM are too small to see. The partial $\chi^2$ value for the ideal estimate is 1.25, and for the bias-corrected estimate it is 203. 
		}
		\label{fig: Phase2NCPAEstimates}
		\vspace{-2mm}
	\end{figure}
	
	The regression equations for the Phase B experiment were set up to jointly estimate the exoplanet image and the NCPA.  
	The NCPA had an RMS of  $0.0495$ radian, which is the uncorrected portion from Phase A.
	The estimated coefficients from each estimator are plotted in Fig.~\ref{fig: Phase2NCPAEstimates}.
	Starting our analysis on the na\"ive estimate, the bias due to the WME is just as noticeable as it was in the exoplanet brightness coefficients, leading to an RMS error in the aberration coefficients of $0.011 \,$radian, which places the error on the same order of magnitude as the NCPA itself.
	With the bias-corrected estimate, we see that where some small residual bias is left even after the correction.
	This is also largely due to the fact that the Monte Carlo wavefronts did not reflect the non-normal character of the AO residual wavefronts.
	In fact, we see that the bias-corrected estimates are actually worse than the na\"ive estimates for several of the coefficients, although the bias-corrected estimates generally are better, as evidenced by an RMS error that is smaller by a factor of about 2.5.
	Looking now to the ideal estimate, which tells us the level of performance that can be reached in an unbiased estimate, the RMS error is smaller than the na\"ive one by a factor of about 62.

	\subsection{The Joint Error Covariance Matrix}
	\label{sssec:ECM}
	
	\begin{figure}[htbp]
		\centering
		\includegraphics[width=.94\linewidth,clip=]{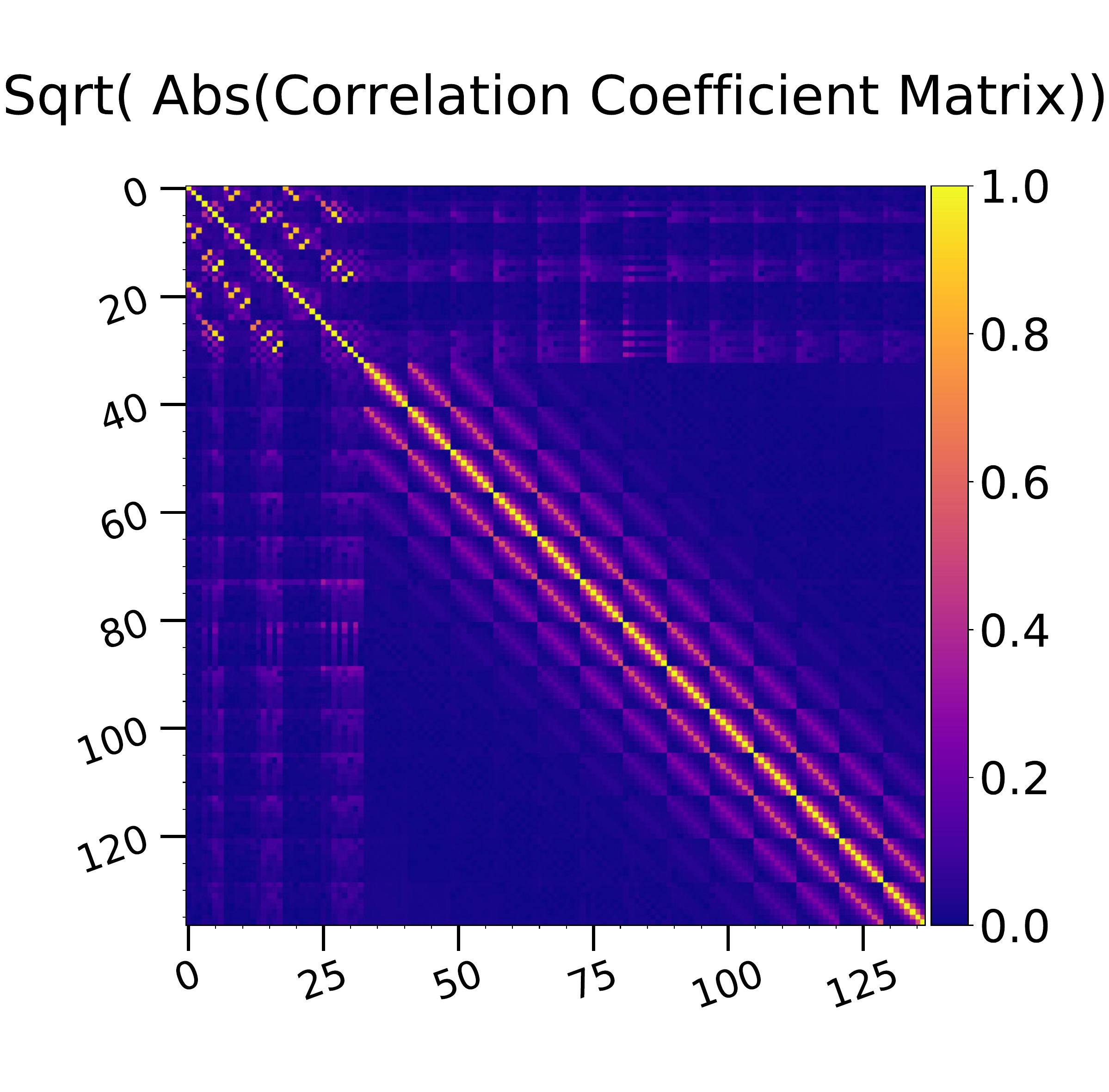}
		\caption{\small The square root of the unsigned correlation coefficient matrix of the ideal regression method for the Phase B experiment. The aberration coefficient coupling reaches a maximum of $0.96$. The coupling for neighboring points on the exoplanet grid is a maximum of about $0.3$. The coupling between the aberration and exoplanet coefficients and reaches a maximum of $0.1$.
		}
		\label{fig: Phase2_JointECMs}
		\vspace{-2mm}
	\end{figure}
	
	We now turn our attention to the interaction between the exoplanet image estimation and the NCPA estimation.
	This can be examined by looking at the error covariance matrix (ECM), $\bC_{\hat{\bx}}$, calculated as part of the regression technique.
	The ideal estimator is simply classical linear regression and its ECM takes a particularly simple form when the weighting matrix of the measurements, is equal to $\bC^{-1}_{\by}$, which is the inverse of error covariance matrix of the noise in the measurements.
	Under this assumption (without this assumption the more complicated formula given in Part II is needed), the ECM, is given by:
	\begin{equation}
	\bC_{\hat{\bx}} = \left (\bA^\rT \bC_{\by}^{-1} \bA \right)^{-1} \, ,
	\label{eq: Error Covariance}
	\end{equation}
	where $^\rT$ indicates matrix transposition and $\bA$\ is the matrix obtained when one vertically stacks all $T$ instances of the matrices $\{ \bA(\bw_t) \}$ in \eqref{eq: y}. 
	The ECM in \eqref{eq: Error Covariance} has two obvious asymptotic (i.e., $T \rightarrow \infty$) behaviors: 1) it is  proportional to $1/T$ and 2) it is proportional to $\bC_{\by}$, which is the covariance of the noise in the science camera measurements.  For small values of $T$, as $T$ increases $\bC_{\bx}$ decreases more quickly than the asymptotic rate because the condition of $\bA$ improves as more rows of $\bA(\bw_t)$ are added.  Indeed, when $T$ is small enough, $\bA$ is likely to be singular, but the random nature of each wavefront, $\bw_t$, is rather helpful in this respect.
	For the bias-corrected estimate, the ECM takes a form that is similar to the classical one, but inflation matrices (calculated with the Monte Carlo wavefronts) must be applied, as explained in Part II.

	One may wonder what happens if the NCPA contains a component that is poorly observed or unobserved by the coronagraph.  
	Firstly, the random modulation of the NCPA by AO residual tends to make it difficult to find modes that are completely unobserved.
	If a mode can be represented by the parameterization of the NCPA in the regression model (i.e., $\theta_p(\ba)$ in \eqref{eq: coronagraph entrance field}), then the error in its estimate can be found from the ECM.
	On the other hand, if the NCPA parameterization cannot fully account for a given mode, one would expect aliasing in the sense that the mode leaks into the parameterized mode, creating an unwanted bias in the estimates of those modes.

	The unsigned correlation coefficient matrix, shown in Fig.~\ref{fig: Phase2_JointECMs}, is defined as $\bC_{\hat{\bx}}' \equiv || \bB \, \bC_{\hat{\bx}} \, \bB  ||$, where $|| \bullet |||$ indicates the absolute value of all matrix elements and $\bB$ is a diagonal matrix chosen so that every element on the diagonal of $\bC_{\hat{\bx}}'$ is unity.
	Without the absolute value operation, $\bC_{\hat{\bx}}'$ would be matrix of correlation coefficients.
	As would be expected for a joint estimate of two types of quantities, we see a four block matrix, with the upper left being the NCPA-NCPA block, the lower right being the exoplanet-exoplanet block, and the other two being NCPA-Exoplanet blocks which are related by a transpose operator since the matrix is real and symmetric.

	Starting with the NCPA-NCPA block (upper left) of the correlation coefficient matrix of the ideal estimate shown in Fig.~\ref{fig: Phase2_JointECMs}, the off diagonal behavior shows of coupling in the estimates of different modes, which is to be expected from the Zernike polynomials used to represent the NCPA in this study.
	There are twenty off-diagonal elements in the upper triangle of this block with values greater than about $0.8$, meaning some the NCPA coefficient estimate errors show strong coupling.
	The exoplanet-exoplanet block (lower right) shows significant power on the diagonal, with the power generally decreasing with distance (the tiling is due to the two-dimensional grid being flattened to a one-dimensional vector).
	The maximum off-diagonal value occurs here for neighboring points on the exoplanet grid, and is about $0.3$.
	The off-diagonal power is largely due to the blurring by the off-axis PSF, which would be worse if the sky-angles in the grid were separated by less than $\lambda / D$, as they are in these simulations.
	It is of course possible to have a more densely sampled grid for the purposes of maximizing the spatial resolution, but we have not yet explored this option in detail.
	
	Next, we come to the NCPA-exoplanet blocks.
	These blocks describe the coupling of the estimate errors in the NCPA and exoplanet coefficients, and show that there is significant coupling.
	The maximum value of the NCPA-exoplanet correlation coefficients is about $0.1$, and we can see that many of the exoplanet coefficient errors have correlations to one or more NCPA coefficients of roughly similar value.
	This tells us that using the joint regression is critical to achieving high contrast in the estimates.

	For an unbiased estimator, the validity of the ECM can then be examined by evaluating the $\chi^2$ test, given by
	\begin{equation}
	\chi^2 = \frac{1}{N_a + N_p}(\hat{\bx} - \bx)^\rT \bC_{\hat{\bx}}^{-1} (\hat{\bx} - \bx) \, ,
	\label{eq: chisquared}
	\end{equation}
	where $\bx$ is the true value of the coefficient vector (so, this metric is only possible to evaluate in simulations),  $N_a+N_p$ is the length of $\bx$, $\hat{\bx}$ is a regression estimate of $\bx$, and $\bC_{\hat{\bx}}$ is the ECM from the regression model.
	With $\chi^2$ so defined, we expect its value to be close to unity if the ECM does indeed correctly characterize the estimate errors.
	The partial and total $\chi^2$ values for the bias-corrected and ideal models can be found in Table (\ref{Table: chi2 values}).
	The partial $\chi^2$ values come from evaluating the components of $\hat{\bx}$ (aberration coefficients or exoplanet brightness coefficients) with the corresponding diagonal block of the ECM.
	The total $\chi^2$ is evaluated using the full $\hat{\bx}$ vector, meaning it also includes the entire ECM (including the off-diagonal blocks).
	As we expect, the total $\chi^2$ value for the ideal estimate is very close to $1$.
	The $\chi^2$ value of about $200$ obtained for the bias-corrected estimate tells us that its ECM is too small.
	We believe that much of this discrepancy is the result of the non-normality of the AO residual wavefronts that was not captured in the Monte Carlo wavefronts used by the bias-corrected estimator.  
	
	The square root of the main diagonal of the ECM can be treated as the error bars (which are too small to be seen in Figs. \ref{fig:EstImagingCoeffs} and \ref{fig: Phase2NCPAEstimates}).
	While the error bars of the ideal estimate are reasonable, as the $\chi^2$ test tells us, the error bars of the bias-corrected estimate are barely any larger and do not come close to describing the errors.
	Again, we attribute this unfortunate circumstance to the fact that our Monte Carlo wavefronts did not account for the non-normality of the true wavefronts.
	It is not unfair to say when the Monte Carlo wavefronts are missing some key elements of realism that the resulting ECM does not describe the errors in the estimates, even when the estimates themselves may be quite accurate and useful.
	Nevertheless, the ECM still tells how the errors in the estimates of various quantities are likely to be coupled.	
	
	\begin{table}[htbp]
		\centering
		\begin{tabular}{| p{2.2cm} | p{1.45cm} | p{1.75cm} | p{1.45cm} |}
			\hline
			\textbf{Estimate Type} & \textbf{$\chi^2$ NCPA} & \textbf{$\chi^2$ Imaging} & \textbf{Total $\chi^2$}\\
			\hline
			Bias-Corrected &  $203$ & $225$ & $232$ \\
			\hline
			Ideal &  $1.25$ & $0.76$ & $0.97$ \\
			\hline
		\end{tabular}
		\caption{\small $\chi^2$ test values for the Phase B experiment (see \eqref{eq: chisquared}). he partial $\chi^2$ for the NCPA and Imaging are computed by only examining the corresponding components of the $\hat{\bx}$ vector and block from the ECM. The total $\chi^2$ is computed using the entire $\hat{\bx}$ vector and ECM.
}		
		\label{Table: chi2 values}
	\end{table}

\section{Conclusions}
\label{sec:Conclusions}

This article presents realistic simulations of regressions based on simultaneous millisecond telemetry from a WFS and science camera behind a stellar coronagraph.
The simulations include self-consistent treatment of an AO system with a pyramid wavefront sensor, a Lyot coronagraph, and photon counting and readout noise in the detectors.  
The objective of the regressions is simultaneous estimation of the non-common path aberrations (NCPA) and the exoplanet image.
We presented two realizable regression models as well as a non-realizable one, which we used as a benchmark.
The two realizable regression models are called the \emph{na\"ive estimator} and the \emph{bias-corrected estimator}, and the non-realizable one is called the \emph{ideal estimator}.
The fact just a few minutes of simulated sky time allowed us to make estimates of the NCPA with RMS errors that were much smaller than the NCPA themselves suggests that our methods could be implemented inside a control loop that compensates for the NCPA in real time.
This is under investigation.
Furthermore, analysis of the error covariance matrix of the estimators demonstrate that the errors in estimating the NCPA and the exoplanet intensity are correlated, suggesting that exoplanet imaging and determination of the NCPA must be done self-consistently to achieve high contrast.

We illustrated the utility of the na\"ive estimator by estimating NCPA of 0.52~radian ($\sim \lambda / 12$) RMS with an accuracy of 0.06~radian ($\sim \lambda/100$) RMS error using only $1$ minute of simulated observation time of an 8\underline{th} magnitude star.
Then, in a followup observation sequence consisting of $4$ minutes of simulated observation time, we assumed an NCPA of 0.05~radian RMS.
We found that the error of the bias-corrected estimate of the NCPA was 0.004~radian ($\sim \lambda/1600$) RMS, while jointly estimating planetary image on a $13\times 8$ object grid (focal plane distances from center ranging from 3-10 $\lambda/D$).
The bias-corrected estimate of the exoplanet scene was nearly identical to the image that would be obtained by PSF subtraction if the PSF were \emph{exactly} known.  
The bias-corrected estimate obtained a $5\, \sigma$ contrast at $3 \, \lambda / D$ of $\sim 1.7 \times10^{-4}$, while at $10 \, \lambda / D$ it was $\sim 2.1 \times 10^{-5}$.
The contrast achieved by the bias-corrected estimator was limited by our inability to draw Monte Carlo samples from non-normal probability density governing the statistics of the AO residual wavefronts.
Part II reports additional simulation results for an experiment in which the AO residuals were drawn from a multivariate normal distribution instead of a simulated AO system.
This simplification allowed us to draw Monte Carlo samples from the distribution as the AO residuals.
In that experiment, the $5 \, \sigma$ contrast achieved by the bias-corrected estimator from  $\sim 17\,$ m of simulated sky time ($T=10^6$) at $3 \, \lambda/D$ was $5.5 \times 10^{-6}$ and at $10 \, \lambda/D$ it was $2.9 \times 10^{-6}$; these values were almost the same as those from the ideal estimate.
The contrasts reported here should  not be interpreted as a fundamental limits to what can be obtained by the regression methods employed, rather they should be seen as illustrative exampled within the context of our simulations.


For comparison, recent efforts using SCExAO with the VAMPIRES module in a bandpass containing the H$\alpha$ line achieved a $5 \, \sigma$ contrast of $\sim 10^{-3}$ at a distance of $\sim 17 \, \lambda/D $ (the Strehl ratio was about $0.45$).\cite{Uyama_etal_2020}
Achieving this contrast required both angular differential imaging (ADI) and spectral differential imaging (SDI) to be applied in post-processing.
As far as we know, the highest contrast ever achieved on-sky was reported in 2015 by Vigan et~al. on the SPHERE AO system.\cite{vigan_etal_2015}
In this case, SPHERE was looking for a second companion to Sirius A (magnitude -1.46), which allowed the AO system to operate at a Strehl in excess of $0.9$ (at $\lambda = 1.6 \, \mu$m).
The 59 minutes of observations were collected over a 2.5 hour period with the coronagraph feeding an integral field spectrograph (IFS) covering the range $\lambda = 0.95$ to $2.3 \, \mu$m.
The field rotation over the 2.5 hour period combined with imaging spectrograph data allowed the authors to apply both ADI and SDI in a post-processing step.
The $5 \, \sigma$ contrast finally achieved was $\sim 5 \times 10^{-5}$ at a distance of $0.2$ arcsec, which corresponds to $3.5$ and $8.5 \, \lambda/D$ at $\lambda = 2.3$ and $0.95 \, \mu$m, respectively. 
At a distance of 0.4 arcsec, they reported a contrast of $\sim 3.5 \times 10^{-6}$.
Thus, SPHERE's reported contrasts for this observation are quite comparable to contrasts achieved in our simulation.
While simulations cannot carry the same weight as results from physical experiments, it is instructive to point out that our simulations in some sense address an imaging problem that is much more difficult than this particular SPHERE experiment:
\begin{itemize}
	\item The SPHERE experiment collected spectral data over more than factor of 2 in wavelength, which allowed SDI, while our data were monochromatic.  (We note that extending our method to multi-wavelength data is relatively straightforward.)
	\item The field rotation during the SPHERE observation allowed ADI, while we had no field rotation. (Including field rotation into our regression equations is easy.)
	\item SPHERE's Strehl ratio was over 0.9, while ours was about 0.75.
	\item Sirius A is brighter than our magnitude 8 source by a factor of over 7000, and SPHERE collected photons for about an hour, whereas we simulated only 4 minutes of sky time.
\end{itemize}
The fact that we obtained such good results without the benefit of multi-wavelength data or field rotation with only a few minutes of simulated sky time speaks to the richness of the millisecond data sets we seek to exploit.
Perhaps it even suggests that our development of the regression models is on a useful track.

	Both the na\"ive and bias-corrected estimators require models of the WFS and coronagraph optical trains, but the bias-corrected estimator also needs knowledge of the spatial statistics of the AO residual wavefronts.
	Model errors and inaccurate characterization of the AO residual statistics will cause unwanted biases in the estimates.
	A closed-form expression for the size of the bias resulting from violating a given assumption is generally not easy to obtain.
	Finding the biases is most easily done for specific cases via simulation studies. 
	In these simulations, we have seen an example of one of the assumptions not holding true: the AO residual wavefronts did not obey multivariate normal statistics, but bias-corrected estimator employed the assumption of a multivariate normal when creating the Monte Carlo wavefronts that it utilizes as part of its machinery (explained in Part II).
	It should be remembered any ground-based exoplanet imaging method must contend with NCPA and the statistical properties of the AO residual wavefronts, so these issues are not unique to our approach.
	
In order to implement this type of regression method on sky, there are number of technical challenges that must be overcome:
\begin{itemize}
	
	\item The regression equations require accurate numerical models of the WFS and coronagraph optical trains.
	These models may contain free parameters representing such things as alignment drifts that are determined from the data, much as the NCPA coefficients are in this study.
	Removing high spatial frequencies from the beam via applying a stop in a focal plane at one or more locations in the apparatus may prove critical for limiting the degrees of freedom that must be taken into account in the regression models.
	
	\item While an initial application may work with a bandpass narrow enough to ignore chromatic effects in the AO residual, moving to larger bandpasses may require taking chromatic considerations into account.
	
	\item This study ignored amplitude effects (scintillation).  
	Ideally, the wavefront sensing scheme would measure amplitudes as well as phase.
	
	\item The Monte Carlo calculations needed by the bias-corrected estimator require the statistics of the AO residual wavefronts.

\end{itemize}

\section*{Acknowledgments}
This work has been supported by the NSF (Awards \#1710356 and \#1600138) and the Heising-Simons Foundation (Grants \#2020-1824 and \#2020-1826). 
We also would like to thank Laurent Jolissaint for his input on the manuscript.
Finally, we would like to thank the anonymous referees for their remarks on improving this work substantially.

\section*{Disclosures}
The authors declare no conflicts of interest.


\begin{thebibliography}{10}
\newcommand{\enquote}[1]{``#1''}

\bibitem{Guyon_habitable_ELT12}
O.~Guyon, F.~Martinache, E.~Cadyd, R.~Belikove, B.~Kunjithapathamd, D.~Wilsond,
  C.~Clergeona, and M.~Mateenc, \enquote{How elts will acquire the first
  spectra of rocky habitable planets,} in \emph{Proc. of SPIE Vol,}  vol. 8447
  (2012), pp. 84471X--1.

\bibitem{Stark_ExoEarthYield14}
C.~C. {Stark}, A.~{Roberge}, A.~{Mandell}, M.~{Clampin}, S.~D.
  {Domagal-Goldman}, M.~W. {McElwain}, and K.~R. {Stapelfeldt}, \enquote{{Lower
  Limits on Aperture Size for an ExoEarth Detecting Coronagraphic Mission},}
  {\protect\JournalTitle{Astrophysical Journal}} \textbf{808}, 149 (2015).

\bibitem{Males_DirectImaging2014}
J.~R. Males, L.~M. Close, O.~Guyon, K.~M. Morzinski, A.~Puglisi, P.~Hinz, K.~B.
  Follette, J.~D. Monnier, V.~Tolls, T.~J. Rodigas \emph{et~al.},
  \enquote{Direct imaging of exoplanets in the habitable zone with adaptive
  optics,} in \emph{SPIE Astronomical Telescopes+ Instrumentation,}
  (International Society for Optics and Photonics, 2014), p. 914820.

\bibitem{Brown_PlanetSearch15}
R.~A. {Brown}, \enquote{{Science Parametrics for Missions to Search for
  Earth-like Exoplanets by Direct Imaging},}
  {\protect\JournalTitle{Astrophysical Journal}} \textbf{799}, 87 (2015).

\bibitem{Guyon_ExAOreview18}
O.~{Guyon}, \enquote{{Extreme Adaptive Optics},} {\protect\JournalTitle{Ann.
  Rev. Astron. Astrophys.}} \textbf{56}, 315--355 (2018).

\bibitem{Digby_Lyot06}
A.~P. Digby, S.~Hinkley, B.~R. Oppenheimer, A.~Sivaramakrishnan, J.~P. Lloyd,
  M.~D. Perrin, J.~Lewis C.~Roberts, R.~Soummer, D.~Brenner, R.~B. Makidon,
  M.~Shara, J.~Kuhn, J.~Graham, P.~Kalas, and L.~Newburgh, \enquote{The
  challenges of coronagraphic astrometry,} {\protect\JournalTitle{The
  Astrophysical Journal}} \textbf{650}, 484--496 (2006).

\bibitem{Racine99}
R.~{Racine}, G.~A.~H. {Walker}, D.~{Nadeau}, R.~{Doyon}, and C.~{Marois},
  \enquote{{Speckle Noise and the Detection of Faint Companions},}
  {\protect\JournalTitle{Pub. Astron. Soc. Pacific}} \textbf{111}, 587--594
  (1999).

\bibitem{Guyon_Limits_ApJ2005}
O.~{Guyon}, \enquote{{Limits of Adaptive Optics for High-Contrast Imaging},}
  {\protect\JournalTitle{Astrophysical Journal}} \textbf{629}, 592--614 (2005).

\bibitem{Boccaletti04}
A.~{Boccaletti}, P.~{Riaud}, P.~{Baudoz}, J.~{Baudrand}, D.~{Rouan},
  D.~{Gratadour}, F.~{Lacombe}, and A.-M. {Lagrange}, \enquote{{The
  Four-Quadrant Phase Mask Coronagraph. IV. First Light at the Very Large
  Telescope},} {\protect\JournalTitle{Pub. Astron. Soc. Pacific}} \textbf{116},
  1061--1071 (2004).

\bibitem{Martinez13}
P.~{Martinez}, M.~{Kasper}, A.~{Costille}, J.~F. {Sauvage}, K.~{Dohlen},
  P.~{Puget}, and J.~L. {Beuzit}, \enquote{{Speckle temporal stability in XAO
  coronagraphic images. II. Refine model for quasi-static speckle temporal
  evolution for VLT/SPHERE},} {\protect\JournalTitle{Astronomy and
  Astrophysics}} \textbf{554}, A41 (2013).

\bibitem{Sparks02}
W.~B. {Sparks} and H.~C. {Ford}, \enquote{{Imaging Spectroscopy for Extrasolar
  Planet Detection},} {\protect\JournalTitle{Astrophysical Journal}}
  \textbf{578}, 543--564 (2002).

\bibitem{Marois00}
C.~{Marois}, R.~{Doyon}, R.~{Racine}, and D.~{Nadeau}, \enquote{{Efficient
  Speckle Noise Attenuation in Faint Companion Imaging},}
  {\protect\JournalTitle{Pub. Astron. Soc. Pacific}} \textbf{112}, 91--96
  (2000).

\bibitem{Pueyo_LOCI12}
L.~{Pueyo}, J.~R. {Crepp}, G.~{Vasisht}, D.~{Brenner}, B.~R. {Oppenheimer},
  N.~{Zimmerman}, S.~{Hinkley}, I.~{Parry}, C.~{Beichman}, L.~{Hillenbrand},
  L.~C. {Roberts}, R.~{Dekany}, M.~{Shao}, R.~{Burruss}, A.~{Bouchez},
  J.~{Roberts}, and R.~{Soummer}, \enquote{{Application of a Damped Locally
  Optimized Combination of Images Method to the Spectral Characterization of
  Faint Companions Using an Integral Field Spectrograph},}
  {\protect\JournalTitle{Astrophysical Journals}} \textbf{199}, 6 (2012).

\bibitem{Rameau_ADI_SDI_limits15}
J.~{Rameau}, G.~{Chauvin}, A.-M. {Lagrange}, A.-L. {Maire}, A.~{Boccaletti},
  and M.~{Bonnefoy}, \enquote{{Detection limits with spectral differential
  imaging data},} {\protect\JournalTitle{Astronomy and Astrophysics}}
  \textbf{581}, A80 (2015).

\bibitem{Marois06}
C.~{Marois}, D.~{Lafreni{\`e}re}, R.~{Doyon}, B.~{Macintosh}, and D.~{Nadeau},
  \enquote{{Angular Differential Imaging: A Powerful High-Contrast Imaging
  Technique},} {\protect\JournalTitle{Astrophysical Journal}} \textbf{641},
  556--564 (2006).

\bibitem{Mawet_SmallAngleReview12}
D.~{Mawet}, L.~{Pueyo}, P.~{Lawson}, L.~{Mugnier}, W.~{Traub}, A.~{Boccaletti},
  J.~T. {Trauger}, S.~{Gladysz}, E.~{Serabyn}, J.~{Milli}, R.~{Belikov},
  M.~{Kasper}, P.~{Baudoz}, B.~{Macintosh}, C.~{Marois}, B.~{Oppenheimer},
  H.~{Barrett}, J.-L. {Beuzit}, N.~{Devaney}, J.~{Girard}, O.~{Guyon},
  J.~{Krist}, B.~{Mennesson}, D.~{Mouillet}, N.~{Murakami}, L.~{Poyneer},
  D.~{Savransky}, C.~{V{\'e}rinaud}, and J.~K. {Wallace}, \enquote{{Review of
  small-angle coronagraphic techniques in the wake of ground-based
  second-generation adaptive optics systems},} in \emph{Society of
  Photo-Optical Instrumentation Engineers (SPIE) Conference Series,}  vol. 8442
  of \emph{Society of Photo-Optical Instrumentation Engineers (SPIE) Conference
  Series} (2012).

\bibitem{Marois_SOSIE}
C.~{Marois}, B.~{Macintosh}, and J.-P. {V{\'e}ran}, \enquote{{Exoplanet imaging
  with LOCI processing: photometry and astrometry with the new SOSIE
  pipeline},} in \emph{Society of Photo-Optical Instrumentation Engineers
  (SPIE) Conference Series,}  vol. 7736 of \emph{Society of Photo-Optical
  Instrumentation Engineers (SPIE) Conference Series} (2010).

\bibitem{Galicher_LOCI}
R.~{Galicher} and C.~{Marois}, \enquote{{Astrometry and photometry in high
  contrast imaging},} in \emph{Second International Conference on Adaptive
  Optics for Extremely Large Telescopes, id.P25,}  (2011), p. 25P.

\bibitem{Hinnen_H2control}
K.~Hinnen, M.~Verhaegen, and N.~Doelman, \enquote{A data-driven $h_2$-optimal
  control approach for adaptive optics,} {\protect\JournalTitle{Control Systems
  Technology, IEEE Transactions on}} \textbf{16}, 381--395 (2008).

\bibitem{Martinache_SpeckleCancel14}
F.~{Martinache}, O.~{Guyon}, N.~{Jovanovic}, C.~{Clergeon}, G.~{Singh},
  T.~{Kudo}, T.~{Currie}, C.~{Thalmann}, M.~{McElwain}, and M.~{Tamura},
  \enquote{{On-Sky Speckle Nulling Demonstration at Small Angular Separation
  with SCExAO},} {\protect\JournalTitle{Pub. Astron. Soc. Pacific}}
  \textbf{126}, 565--572 (2014).

\bibitem{Martinache_ClosedLoop16}
F.~{Martinache}, N.~{Jovanovic}, and O.~{Guyon}, \enquote{{Closed-loop focal
  plane wavefront control with the SCExAO instrument},}
  {\protect\JournalTitle{Astronomy and Astrophysics}} \textbf{593}, A33 (2016).

\bibitem{Matthews_EFC_P1640_17}
C.~T. Matthews, J.~R. Crepp, G.~Vasisht, and E.~Cady, \enquote{Electric field
  conjugation for ground-based high-contrast imaging: robustness study and
  tests with the project 1640 coronagraph,} {\protect\JournalTitle{Journal of
  Astronomical Telescopes, Instruments, and Systems}} \textbf{3}, 1 -- 12 -- 12
  (2017).

\bibitem{jovanovic2018review}
N.~Jovanovic, O.~Absil, P.~Baudoz, M.~Beaulieu, M.~Bottom, E.~Cady,
  B.~Carlomagno, A.~Carlotti, D.~Doelman, K.~Fogarty, R.~Galicher, O.~Guyon,
  S.~Haffert, E.~Huby, J.~Jewell, C.~Keller, M.~A. Kenworthy, J.~Knight,
  J.~Kuhn, K.~Miller, J.~Mazoyer, M.~N'Diaye, E.~Por, L.~Pueyo, A.~J.~E. Riggs,
  G.~Ruane, D.~Sirbu, F.~Snik, J.~K. Wallace, M.~Wilby, and M.~Ygouf,
  \enquote{Review of high-contrast imaging systems for current and future
  ground-based and space-based telescopes ii. common path wavefront
  sensing/control and coherent differential imaging,}  (2018).

\bibitem{Kasdin_EFC13}
T.~D. Groff and N.~J. Kasdin, \enquote{Kalman filtering techniques for focal
  plane electric field estimation,} {\protect\JournalTitle{J. Opt. Soc. Am. A}}
  \textbf{30}, 128--139 (2013).

\bibitem{Potier_2020}
A.~Potier, P.~Baudoz, R.~Galicher, G.~Singh, and A.~Boccaletti,
  \enquote{Comparing focal plane wavefront control techniques: Numerical
  simulations and laboratory experiments,} {\protect\JournalTitle{Astronomy and
  Astrophysics}} \textbf{635}, A192 (2020).

\bibitem{Thomas_GroundEFC10}
S.~J. {Thomas}, A.~A. {Give'On}, D.~{Dillon}, B.~{Macintosh}, D.~{Gavel}, and
  R.~{Soummer}, \enquote{{Laboratory test of application of electric field
  conjugation image-sharpening to ground-based adaptive optics},} in
  \emph{Society of Photo-Optical Instrumentation Engineers (SPIE) Conference
  Series,}  vol. 7736 of \emph{Society of Photo-Optical Instrumentation
  Engineers (SPIE) Conference Series} (2010).

\bibitem{Vigan_ZELDA_NCPA_AA19}
A.~{Vigan}, M.~{N'Diaye}, K.~{Dohlen}, J.~F. {Sauvage}, J.~{Milli}, G.~{Zins},
  C.~{Petit}, Z.~{Wahhaj}, F.~{Cantalloube}, A.~{Caillat}, A.~{Costille},
  J.~{Le Merrer}, A.~{Carlotti}, J.~L. {Beuzit}, and D.~{Mouillet},
  \enquote{{Calibration of quasi-static aberrations in exoplanet direct-imaging
  instruments with a Zernike phase-mask sensor. III. On-sky validation in
  VLT/SPHERE},} {\protect\JournalTitle{Astronomy and Astrophysics}}
  \textbf{629}, A11 (2019).

\bibitem{Paul_PhaseDivers13}
B.~{Paul}, J.-F. {Sauvage}, and L.~M. {Mugnier}, \enquote{{Coronagraphic phase
  diversity: performance study and laboratory demonstration},}
  {\protect\JournalTitle{Astronomy and Astrophysics}} \textbf{552}, A48 (2013).

\bibitem{Schiller_COFFEE19}
O.~{Herscovici-Schiller}, J.-F. {Sauvage}, L.~M. {Mugnier}, K.~{Dohlen}, and
  A.~{Vigan}, \enquote{{Coronagraphic phase diversity through residual
  turbulence: performance study and experimental validation},}
  {\protect\JournalTitle{Monthly Notices Royal Astron. Soc.}} \textbf{488},
  4307--4316 (2019).

\bibitem{Saphira_eAPD14}
G.~{Finger}, I.~{Baker}, D.~{Alvarez}, D.~{Ives}, L.~{Mehrgan}, M.~{Meyer},
  J.~{Stegmeier}, and H.~J. {Weller}, \enquote{{SAPHIRA detector for infrared
  wavefront sensing},} in \emph{Society of Photo-Optical Instrumentation
  Engineers (SPIE) Conference Series,}  vol. 9148 of \emph{Society of
  Photo-Optical Instrumentation Engineers (SPIE) Conference Series} (2014),
  p.~17.

\bibitem{Mazin_MKIDS18}
S.~R. {Meeker}, B.~A. {Mazin}, A.~B. {Walter}, P.~{Strader}, N.~{Fruitwala},
  C.~{Bockstiegel}, P.~{Szypryt}, G.~{Ulbricht}, G.~{Coiffard}, B.~{Bumble},
  G.~{Cancelo}, T.~{Zmuda}, K.~{Treptow}, N.~{Wilcer}, G.~{Collura},
  R.~{Dodkins}, I.~{Lipartito}, N.~{Zobrist}, M.~{Bottom}, J.~C. {Shelton},
  D.~{Mawet}, J.~C. {van Eyken}, G.~{Vasisht}, and E.~{Serabyn},
  \enquote{{DARKNESS: A Microwave Kinetic Inductance Detector Integral Field
  Spectrograph for High-contrast Astronomy},} {\protect\JournalTitle{Pub.
  Astron. Soc. Pacific}} \textbf{130}, 065001 (2018).

\bibitem{Frazin13}
R.~A. {Frazin}, \enquote{{Utilization of the Wavefront Sensor and
  Short-exposure Images for Simultaneous Estimation of Quasi-static Aberration
  and Exoplanet Intensity},} {\protect\JournalTitle{Astrophysical Journal}}
  \textbf{767}, 21 (2013).

\bibitem{Fitz_SpeckleStats06}
M.~P. {Fitzgerald} and J.~R. {Graham}, \enquote{{Speckle Statistics in
  Adaptively Corrected Images},} {\protect\JournalTitle{Astrophysical Journal}}
  \textbf{637}, 541--547 (2006).

\bibitem{Gladysz10}
S.~{Gladysz}, N.~{Yaitskova}, and J.~C. {Christou}, \enquote{{Statistics of
  intensity in adaptive-optics images and their usefulness for detection and
  photometry of exoplanets},} {\protect\JournalTitle{Journal of the Optical
  Society of America A}} \textbf{27}, A260000--A75 (2010).

\bibitem{Stangalini_RecurQuant18}
M.~{Stangalini}, G.~{Li Causi}, F.~{Pedichini}, S.~{Antoniucci}, M.~{Mattioli},
  J.~{Christou}, G.~{Consolini}, D.~{Hope}, S.~M. {Jefferies}, and
  R.~{Piazzesi}, \enquote{{Recurrence Quantification Analysis as a
  Post-processing Technique in Adaptive Optics High-contrast Imaging},}
  {\protect\JournalTitle{Astrophysical Journal}} \textbf{868}, 6 (2018).

\bibitem{Walter_ModifiedRician19}
A.~B. {Walter}, C.~{Bockstiegel}, T.~D. {Brandt}, and B.~A. {Mazin},
  \enquote{{Stochastic Speckle Discrimination with Time-tagged Photon Lists:
  Digging below the Speckle Noise Floor},} {\protect\JournalTitle{Pub. Astron.
  Soc. Pacific}} \textbf{131}, 114506 (2019).

\bibitem{Sauvage_model10}
J.-F. Sauvage, L.~M. Mugnier, G.~Rousset, and T.~Fusco, \enquote{Analytical
  expression of long-exposure adaptive-optics-corrected coronagraphic image.
  first application to exoplanet detection,} {\protect\JournalTitle{J. Opt.
  Soc. Am. A}} \textbf{27}, A157--A170 (2010).

\bibitem{Codona13}
J.~L. {Codona} and M.~{Kenworthy}, \enquote{{Focal Plane Wavefront Sensing
  Using Residual Adaptive Optics Speckles},}
  {\protect\JournalTitle{Astrophysical Journal}} \textbf{767}, 100 (2013).

\bibitem{Fusco_AOdecon99}
T.~{Fusco}, J.~P. {V{\'e}ran}, J.~M. {Conan}, and L.~M. {Mugnier},
  \enquote{{Myopic deconvolution method for adaptive optics images of stellar
  fields},} {\protect\JournalTitle{Astron. and Astrophys. Supp.}} \textbf{134},
  193--200 (1999).

\bibitem{Jefferies_WFSdecon13}
Q.~Chu, S.~Jefferies, and J.~G. Nagy, \enquote{Iterative wavefront
  reconstruction for astronomical imaging,} {\protect\JournalTitle{SIAM Journal
  on Scientific Computing}} \textbf{35}, S84--S103 (2013).

\bibitem{Schulz_HubbleBDC97}
T.~J. {Schulz}, B.~E. {Stribling}, and J.~J. {Miller}, \enquote{{Multiframe
  blind deconvolution with real data: imagery of the Hubble Space Telescope},}
  {\protect\JournalTitle{Optics Express}} \textbf{1}, 355 (1997).

\bibitem{Fetick_BDC20}
R.~{F{\'e}tick}, L.~{Mugnier}, T.~{Fusco}, and B.~{Neichel}, \enquote{{Blind
  deconvolution in astronomy with adaptive optics: the parametric marginal
  approach},} {\protect\JournalTitle{arXiv e-prints}} arXiv:2006.11160 (2020).

\bibitem{IntroFourierOptics}
J.~W. {Goodman}, \emph{Introduction to Fourier Optics, second edition} (The
  McGraw-Hill Companies, Inc., 1996).

\bibitem{Frazin_JOSAA2018}
R.~A. {Frazin}, \enquote{{Efficient, nonlinear phase estimation with the
  nonmodulated pyramid wavefront sensor},} {\protect\JournalTitle{Journal of
  the Optical Society of America A}} \textbf{35}, 594 (2018).

\bibitem{StatisticalOptics}
J.~W. {Goodman}, \emph{Statistical Optics, 2nd Edition} (John Wiley and Sons,
  Inc., 2015).

\bibitem{Assemat_Wilson_InfinitePhaseScreen06}
F.~Ass\'{e}mat, R.~W. Wilson, and E.~Gendron, \enquote{Method for simulating
  infinitely long and non stationary phase screens with optimized memory
  storage,} {\protect\JournalTitle{Opt. Express}} \textbf{14}, 988--999 (2006).

\bibitem{Por_Haffert_Hcipy2018}
E.~H. Por, S.~Y. Haffert, V.~M. Radhakrishnan, D.~S. Doelman, M.~van Kooten,
  and S.~P. Bos, \enquote{{High Contrast Imaging for Python (HCIPy): an
  open-source adaptive optics and coronagraph simulator},} in \emph{Adaptive
  Optics Systems VI,}  vol. 10703 L.~M. Close, L.~Schreiber, and D.~Schmidt,
  eds., International Society for Optics and Photonics (SPIE, 2018), pp. 1112
  -- 1125.

\bibitem{2018SPIE10703E..09M}
J.~R. {Males}, L.~M. {Close}, K.~{Miller}, L.~{Schatz}, D.~{Doelman},
  J.~{Lumbres}, F.~{Snik}, A.~{Rodack}, J.~{Knight}, K.~{Van Gorkom}, J.~D.
  {Long}, A.~{Hedglen}, M.~{Kautz}, N.~{Jovanovic}, K.~{Morzinski}, O.~{Guyon},
  E.~{Douglas}, K.~B. {Follette}, J.~{Lozi}, C.~{Bohlman}, O.~{Durney},
  V.~{Gasho}, P.~{Hinz}, M.~{Ireland}, M.~{Jean}, C.~{Keller}, M.~{Kenworthy},
  B.~{Mazin}, J.~{Noenickx}, D.~{Alfred}, K.~{Perez}, A.~{Sanchez}, C.~{Sauve},
  A.~{Weinberger}, and A.~{Conrad}, \enquote{{MagAO-X: project status and first
  laboratory results},} in \emph{Proc. SPIE,}  vol. 10703 (2018), p. 1070309.

\bibitem{Lozi_SCExAO_2018}
J.~Lozi and {others}, \enquote{{SCExAO}, an instrument with a dual purpose:
  perform cutting-edge science and develop new technologies,} in \emph{Adaptive
  {Optics} {System} {VI},}  vol. 10703-270 of \emph{Proc. {SPIE}} (2018).

\bibitem{Herscovici-Schiller2017}
O.~{Herscovici-Schiller}, L.~M. {Mugnier}, and J.-F. {Sauvage}, \enquote{{An
  analytic expression for coronagraphic imaging through turbulence. Application
  to on-sky coronagraphic phase diversity},} {\protect\JournalTitle{Monthly
  Notices Royal Astron. Soc.}} \textbf{467}, L105--L109 (2017).

\bibitem{Guyon_LyotPIAA14}
O.~{Guyon}, P.~M. {Hinz}, E.~{Cady}, R.~{Belikov}, and F.~{Martinache},
  \enquote{{High Performance Lyot and PIAA Coronagraphy for Arbitrarily Shaped
  Telescope Apertures},} {\protect\JournalTitle{Astrophysical Journal}}
  \textbf{780}, 171 (2014).

\bibitem{Por_APPdesign_2017}
E.~H. Por, \enquote{{Optimal design of apodizing phase plate coronagraphs},} in
  \emph{Techniques and Instrumentation for Detection of Exoplanets VIII,}  vol.
  10400 S.~Shaklan, ed., International Society for Optics and Photonics (SPIE,
  2017), pp. 236 -- 247.

\bibitem{Otten_vAPPCperformance14}
G.~P. P.~L. Otten, F.~Snik, M.~A. Kenworthy, M.~N. Miskiewicz, and M.~J.
  Escuti, \enquote{Performance characterization of a broadband vector apodizing
  phase plate coronagraph,} {\protect\JournalTitle{Opt. Express}} \textbf{22},
  30287--30314 (2014).

\bibitem{Mawet_VectorVortex2009}
D.~Mawet, J.~T. Trauger, E.~Serabyn, D.~C. Moody, K.~M. Liewer, J.~E. Krist,
  D.~M. Shemo, and N.~A. O'Brien, \enquote{{Vector vortex coronagraph: first
  results in the visible},} in \emph{Techniques and Instrumentation for
  Detection of Exoplanets IV,}  vol. 7440 S.~B. Shaklan, ed., International
  Society for Optics and Photonics (SPIE, 2009), pp. 301 -- 310.

\bibitem{MalesandGuyon2018}
J.~R. Males and O.~Guyon, \enquote{{Ground-based adaptive optics coronagraphic
  performance under closed-loop predictive control},}
  {\protect\JournalTitle{Journal of Astronomical Telescopes, Instruments, and
  Systems}} \textbf{4}, 1 -- 21 (2018).

\bibitem{Close2018_MagAOXdesign}
L.~M. Close, J.~R. Males, O.~Durney, C.~Sauve, M.~Kautz, A.~Hedglen, L.~Schatz,
  J.~Lumbres, K.~Miller, K.~V. Gorkom, M.~Jean, and V.~Gasho, \enquote{{Optical
  and mechanical design of the extreme AO coronagraphic instrument MagAO-X},}
  in \emph{Adaptive Optics Systems VI,}  vol. 10703 L.~M. Close, L.~Schreiber,
  and D.~Schmidt, eds., International Society for Optics and Photonics (SPIE,
  2018), pp. 1227 -- 1236.

\bibitem{Uyama_etal_2020}
T.~Uyama, B.~Norris, N.~Jovanovic, J.~Lozi, P.~G. Tuthill, O.~Guyon, T.~Kudo,
  J.~Hashimoto, M.~Tamura, and F.~Martinache, \enquote{{High-contrast H$\alpha$
  imaging with Subaru/SCExAO + VAMPIRES},} {\protect\JournalTitle{Journal of
  Astronomical Telescopes, Instruments, and Systems}} \textbf{6}, 1 -- 17
  (2020).

\bibitem{vigan_etal_2015}
A.~Vigan, C.~Gry, G.~Salter, D.~Mesa, D.~Homeier, C.~Moutou, and F.~Allard,
  \enquote{{High-contrast imaging of Sirius A with VLT/SPHERE: looking for
  giant planets down to one astronomical unit},} {\protect\JournalTitle{Monthly
  Notices of the Royal Astronomical Society}} \textbf{454}, 129--143 (2015).

\end{thebibliography}

\end{document}